\documentclass{aa}  

\usepackage{graphicx}
\usepackage{txfonts}
\usepackage{xcolor}
\usepackage{ulem}
\usepackage{multicol}
\usepackage{mathtools}
\usepackage{amsmath}
\usepackage[toc]{appendix}
\newcommand{\afe}{[$\alpha$/Fe]}

\newcommand{\feh}{[Fe/H]}
\newcommand{\ms}{M$_{\odot}$}

\newcommand{\teff}{$T_\mathrm{eff}$}
\newcommand{\logg}{log~$g$}

\begin{document} 

   \title{Evolution of lithium in the disc of the Galaxy and the role of novae}

   \author{Sviatoslav Borisov
          \inst{1},
          Nikos Prantzos
          \inst{2},
          Corinne Charbonnel
          \inst{1,3}
          }

    \institute{Department of Astronomy, University of Geneva, Chemin Pegasi~51, 1290 Versoix, Switzerland\\
            \email{sviatoslav.borisov@unige.ch}
            \and
            Institut d’Astrophysique de Paris, UMR7095 CNRS, Sorbonne Universit\'{e}, 98bis Bd. Arago, 75104 Paris, France
            \and
            IRAP, CNRS UMR 5277 \& Universit\'{e} de Toulouse, 14 avenue Edouard Belin, 31400 Toulouse, France
            }


 
  \abstract
   {Lithium plays a crucial role in probing stellar physics and stellar and primordial nucleosynthesis, as well as the chemical evolution of our Galaxy. Stars are considered to be the main source of Li, still the identity of its primary stellar producer has long been a matter of debate.}
   {In light of recent theoretical and observational results,  we investigate in this study the role of two candidate sources of Li enrichment in the Milky Way, namely AGB stars and, in particular, novae. }
   {We utilize a one-zone Galactic chemical evolution model to assess the viability of AGB stars and novae as stellar sources of Li. We use recent theoretical Li yields for AGB stars, while for novae we adopt observationally inferred Li yields and recently derived Delay Time Distributions (DTDs). Subsequently, we extend our analysis using a multi-zone model with radial migration to investigate spatial variations in the evolution of Li across the Milky Way disc and compare the results with observational data for field stars and open clusters.}
   {Our analysis shows that AGB stars fail by far to reproduce the meteoritic Li abundance. In contrast, novae appear as promising candidates within the adopted framework, allowing us to quantify the contribution of each Li source at Sun's formation and today. Our multi-zone model reveals the role of the differences in the DTDs of SN~Ia and novae in shaping the evolution of Li in the various galactic zones.   Its results are in fair agreement  with the observational data for most open clusters, but small discrepancies appear in the outer disc.}
   {}


   \maketitle
%

\section{Introduction}
\label{sec:intro}

Lithium is certainly the most intriguing of the chemical elements. Being the third lightest element in the periodic table, it holds a unique position in astrophysics owing to its pivotal role in probing the chemical evolution of the Milky Way and the properties of the outer stellar layers. Despite the relative simplicity of its atomic structure, the abundance and distribution of lithium in stellar environments remain enigmatic and continue to challenge our understanding of stellar structure, nucleosynthesis and galactic chemical evolution. The most abundant isotope of lithium, $^7$Li, is the only nuclide known to be produced in three different astrophysical sites: in the hot early Universe by Big Bang nucleosynthesis (BBN), in Galactic cosmic rays (GCR) through spallation/fusion reactions, and in stars by thermonuclear reactions. The contributions to the total Li abundance of the first two sources can be easily quantified and is found to be less than half of the proto-solar Li value \citep[e.g.,][]{Prantzos2012}. At the same time, the type of stars that make the main ``stellar'' contribution to the Li content of the Galaxy remains a matter of debate.

The production mechanism of the light elements Li, Be and B was an enigma to the founding fathers of stellar nucleosynthesis who coined the name ``x-process'' to it \citep{B2FH_1957}.
The pioneering studies of \citet{Reeves_1970} and \citet{Meneguzzi_1971} showed that the ``x-process'' is due to GCR interacting with the interstellar medium (ISM) and can produce the totality of proto-solar (meteoritic) $^6$Li, $^9$Be, $^{10}$B and most of  $^{11}$B, but only $\sim$20\% of  $^7$Li. 

\cite{Cameron1971} suggested that ``...potentially very large amounts of $^7$Li may be produced by the $^7$Be transport mechanism of \cite{Cameron_1955}'' in red giant stars. Subsequent theoretical   \citep{Scalo_1975,  Sackmann1992, 1997A&AS..123..241F}  and observational \citep{Smith1989,Smith1990} studies showed that this could occur in Asymptotic Giant Branch (AGB) stars: in the ``hot bottom'' (temperatures T$\sim$$(40-60)\times10^6$ K) of their convective envelopes the fusion of $^3$He with $^4$He produces $^7$Be which is convectively transported to the surface while beta-decaying in a timescale of 53 days to $^7$Li. 

In the meantime, \cite{Spite1982} and \cite{Spite2_1984} found that the Li abundances in dwarf halo stars of Pop II display a remarkably constant value for various temperatures (unlike in Pop I stars), and should thus reflect the original Li content of those stars; moreover, the lithium abundance in the hotter halo dwarfs appeared very uniform (independent of metal abundance, spatial velocity, eccentricity and galactic stellar orbit), arguing in favour of the Big Bang origin of lithium, as  calculated by \cite{Wagoner_1967}. Up to a metallicity \feh$\sim$-1.2 \citep[see Fig. 3 in][]{Rebolo_1988}, the "Spite plateau" abundance of Li is $\sim$1/10th of the meteoritic one and, combined with the known contribution of the GCRs,  suggests that the dominant source of Li is the stellar one. 

Beyond AGBs, other candidate stellar sources of Li, are: 

- novae \citep{Arnould_1975,Starrfield_1978,Jose1998}, where explosive H-burning occurs episodically at the surface of a white dwarf accreting material from a companion star; 

- core-collapse supernovae (CCSN), where energetic $\mu$- and $\tau$- neutrinos from the core excite the He nuclei of the outer layers and induce reactions between them and $^3$He or $^3$H leading to the production of $^7$Li \citep{Woosley_1990};

- low-mass red giant stars, via extra deep mixing and the associated ``cool bottom processing'', with the amount of Li produced depending critically on the details of the extra mixing mechanism \citep{Sackmann1999}.

The Galactic chemical evolution of Li was studied mostly with one-zone models, adopting one or all of the aforementioned stellar sources and considering the ``Spite-plateau'' value of Li as the primordial one \citep[e.g.,][]{Abia1988,  Abia1993, Abia1995,DAntona_1991,Matteucci_1995, Travaglio2001,Romano2001,Prantzos2012}. No robust conclusions could be reached as to the main stellar source, but it was shown that the yields of all candidate sources are insufficient to explain the proto-solar Li by a large factor \citep[see][and references therein]{Prantzos2012}. Moreover,  the observational constraints are weak, since Li is prone to depletion in stellar envelopes and its photospheric abundance does not trace the initial one, except in A-type stars \citep[see][and references therein]{Charbonnel2021,Randich_2021}. The  ``upper envelope'' of the Li observations can be used, in principle, as tracer of the true Li evolution \citep{Rebolo_1988} or at least as a lower limit describing that evolution, but it is difficult to define it in practice \citep{Lambert_2004}.

In the past years, novae attracted considerable attention as a source of Li \citep{Cescutti2019,Grisoni2019,Matteucci_2021,Romano2021}, due to the detection of substantial abundances of $^7$Be and, perhaps, $^7$Li in their ejecta \citep[][and references therein]{Tajitsu2015,Izzo_2015,Molaro_2016,Izzo_2018,Molaro2023}. On the other hand, the recent work of \citet{Kemp_phd} provided for the first time theoretical Delay Time Distributions (DTDs) for novae, through population synthesis models of binary star evolution. \citet{Kemp_2022b} used those DTDs and observationally inferred Li yields to study the Li evolution with a one-zone model and to show that novae can be the only stellar source of Li.

In light of those recent developments, we reassess in this work the problem of the Galactic evolution of Li and we compare our results to observations of field stars from the GALAH survey and of open clusters from the \textit{Gaia}-ESO survey. In Sec.~\ref{sec:observations} we present the observational data.  In Sec.~\ref{sec:one-zone} we first discuss in some depth the nova  DTDs (which are metallicity dependent) and the corresponding Li yields. Then we use a one-zone model to show that AGBs are negligible contributors to Li with current Li yields, while novae can be the sole stellar source; we also evaluate the contribution of BBN, GCRs, and novae to the Galactic Li content. In Sec.~\ref{sec:multi-zone} we use a multi-zone model with radial migration to discuss the impact of the adopted nova DTDs on the evolution of Li in the various Galactic zones, paying particular attention to the metal-rich stars formed in the inner disc. We summarize our results in Sec.~\ref{sec:conclusion}.

\section{Observational data}
\label{sec:observations}

\subsection{Upper lithium envelope}
\label{subsec:upper_envelope}

In exploring the chemical evolution of the Milky Way, it is crucial to take into account the evolution of Li in the stars that are used to constrain the models. The abundance of this fragile element is subject to significant depletion within stars, and stars rarely exhibit in their photospheres the Li content they inherited from their protostellar cloud \citep[see][and references therein]{Borisov2024}. To track the original abundance of Li in stars in the Galaxy, we need to turn our attention to the stars that might preserve it as close as possible to the initial content, and the best candidates are warm stars on the hot side of the so-called Li-dip with the effective temperature above $\sim$ 6800~K \citep{Randich2020,Charbonnel2021}.  These hot and relatively massive stars (masses above $\sim 1.3$M$_{\odot}$ at solar metallicity) have shallow convective envelopes, which prevents the transport of Li to the inner hot layers where it can be destroyed by the proton capture reaction $^7$Li(p,$\alpha$)$\alpha$. Unfortunately, these stars are less numerous than G-type stars due to the initial mass function (IMF), and they are also rarely included in surveys, as they are fast rotators for which it is more difficult to determine abundances. 

Unfortunately, the selection of hot dwarfs is complicated by the fact that at lower metallicities such stars are scarce or simply absent, because of the strong dependence of their lifetimes on mass and metallicity, with more metal-poor stars being older, on average, than metal-rich stars. At low metallicity ([Fe/H] below $\sim$ -1.5~dex), we are thus left with the main sequence or turnoff stars that have undergone significant Li depletion along their evolution. For this reason, we expect an increase in the difference in A(Li) between the predictions of Galactic evolution models and the observed stellar Li values, especially at low metallicities.

\subsection{Selection of the sample}
\label{subsec:selection}

We selected dwarf field stars from GALAH~DR3 \citep{Buder2021} in a wide metallicity range from $-2.84\leq$[Fe/H]$\leq 0.5$ (lower limit on metallicity corresponds to the most Fe-poor star in GALAH~DR3). We applied the quality flag on stellar parameters \texttt{flag\_sp}=0 (for \teff \ and \logg). When set to zero, it also eliminates possible binaries. In addition, we applied a flag on iron and lithium abundance: \texttt{flag\_fe\_h}=0 and \texttt{flag\_li\_fe}=0. As recommended by \citet{Buder2021}, we selected stars with S/N$\geq30$. We keep only those objects that have a relative uncertainty of age $\sigma_{age}$/age$\leq$10\% and low interstellar extinction $A_G\leq0.2$. The latter allows us to avoid the influence of interstellar extinction on the \teff \ and $L$ determination, which leads to incorrect age determination.

Finally, to focus on dwarf stars and avoid possible contamination from (sub)giants, we made the preliminary selection of stars with \logg$\geq$3.5 and then followed a similar selection algorithm as in \citet[][see their Sec.~4.1.3]{Borisov2024}. Briefly recapitulate the basic mechanism of the algorithm: we use a grid of stellar evolutionary tracks with the determined positions of the ZAMS and TAMS (terminal-age main sequence) in the Hertzsprung-Russell Diagram (HRD) for each track. For each chemical composition of the grid, we define a polygon with ZAMS and TAMS points as its vertices. We assume that a star is a dwarf if its position in the HRD (\teff \ and \textit{Gaia}~DR3-based luminosity $L$) falls into this polygon. To take into account the discreteness of the grid, we perform bicubic interpolation of vertices on a star's \feh \ and \afe. Here we used the BaSTI tracks \citep{Pietrinferni2004,Pietrinferni2006,Pietrinferni2024} as they have wide coverage in \teff \ space. As a result, the final sample of field stars contains 2136 objects with 3D,NLTE Li abundances. The hottest stars of the sample are on the hot side of the Li-dip and correspond to the ``warm group'' referred to in \citet{Gao2020}. We used the ages of stars provided in the GALAH value-added catalogue and determined with the code BSTEP by \citet{Sharma2018} using \texttt{PARSEC+COLIBRI} isochrones \citep{Marigo2017}. The wide age coverage of the GALAH sample, which ranges from 25~Myr to 12.8~Gyr, allows us to challenge the predictions of our Galactic evolution models over almost the entire history of the MW. 

The vast majority of the sample field stars are from the solar vicinity: 95\% are within 1~kpc from the Sun while the median distance is $\sim$590~pc. This does not allow us to compare our multi-zone model predictions (see Sec.~\ref{sec:multi-zone}) with observations at different galactocentric distances $R_{GC}$. For this reason, we also selected open clusters from \citet{Romano2021} that cover a wide range of $R_{GC}$ from 5.2 to 15~kpc. In the paper, the authors provide mean values of A(Li) (not corrected for non-LTE effects) for the stars that have suffered minimal Li depletion. In most cases (especially in the case of clusters older than 0.1~Gyr), that corresponds to stars from the hot side of the Li-dip that are, on average, slightly hotter than the Li-richest stars from the GALAH sample. Otherwise, the measurements are based on the Li content of pre-main sequence (PMS) stars.

\section{One-zone chemical evolution model for the solar neighbourhood}
\label{sec:one-zone}

\subsection{Description of the model}
\label{subsec:desciption_one-zone}

Our simple, one-zone model, is introduced in detail in \cite{Prantzos2018} and here we present its main features. The local disc is built by infall of gas at an exponentially decreasing rate and a characteristic time-scale of 8~Gyr, where the star formation rate (SFR) $\Psi$ is given by a Schmidt-Kennicutt law:
\begin{equation}
\Psi(t) \ = C \ \Sigma_G(t),
\end{equation}
where $\Sigma_G$ is the local gas surface density and the coefficient $C$ is chosen to obtain a gas fraction of $\sim$20\% at the end of the simulation, compatible with the one presently evaluated in the solar neighbourhood. We assume that the duration of the system is 12~Gyr.

We use the metallicity-dependent stellar lifetimes and yields of \citet{Cristallo2015} for Low and Intermediate Mass stars (1-7~M$_{\odot}$) and of \citet{Limongi2018} for the massive ones (M$>$13M$_{\odot}$), which include mass loss and rotation. No Li is produced in those models (see Sec. 3.3). Stars above 25\ms \ are assumed to collapse into black holes and enrich the interstellar medium only with their wind ejecta. We interpolate the yields in the region 7-13 M$_{\odot}$. For the massive stars, we use an Initial Distribution of Rotational Velocities (IDROV), calibrated on observations of abundances in halo and disc stars \cite[see][for details]{Prantzos2018}. We adopt the stellar IMF of \cite{Kroupa2001} in the mass range 0.1-120~\ms \ and we calculate chemical evolution with the Single Particle Population (SSP) method.

We adopt metallicity-dependent yields of SN~Ia from \cite{Leung2018} and we consider the Delay Time Distribution (DTD) of SN~Ia as discussed in \citet{Maoz2017}, i.e. a power-law in time with index x=-1.1. We impose a lower cut-off at 40~Myr which corresponds to the shortest possible time between the formation of a binary system and the formation of a white dwarf in it. At a given moment $t$, we compute the SN~Ia rate as a convolution of the SFR $\Psi(t)$ and the SN~Ia DTD:  

\begin{equation}
    R_\mathrm{Ia}(t)=\int_0^t\mathrm{DTD}(t') \Psi(t-t') dt'
    \label{eq:rate_snia}
\end{equation}
It should be noted that the one-zone model presented here is developed only for the disc of the Milky Way. Despite this fact, it can serve as a rough approximation for capturing the general evolutionary trends in the Galactic halo, since the halo phase is characterized by metallicities \feh$\leq$-1, a duration of ~1 Gyr and a ~constant \afe \ ratio during that period. These features are also met in the 1-zone model of \citet{Prantzos2018} adopted here.

\subsection{Production of Li: GCRs and nova yields and DTDs}
\label{subsec:nova_treatment}

We model the evolution of Li starting with an initial abundance of A(Li)=2.75~dex  that corresponds to the SBBN value of \citet{Pitrou2018}. This Li is injected into the system during the evolution through the infall of primordial composition. 
 
We compute Li production by GCR as discussed in detail in \cite{Prantzos2012}, in a fully self-consistent manner:  a) we consider the energy released by supernovae (10 \% of the kinetic energy of  $1.5\times10^{51}$~erg from both CCSN and SN~Ia); b) we assume that this energy is distributed across an injection spectrum of GCR (power-law in momentum space) accelerated near each SN;  c) we consider the GCR spectra at equilibrium after injection spectrum is propagated with an escape length $\Lambda$=10~gr/cm$^2$,  4) we take into account the composition of the GCR as resulting from the composition of the stellar winds in the case of CCSN (properly weighted with the adopted IMF) and from the composition of the ISM in the case of SN~Ia; 5) the GCR particles of H-He-C-N-O with the aforementioned equilibrium spectra and composition interact with the ISM H-He-C-N-O with energy-dependent cross-sections and produce the isotopes $^6$Li, $^7$Li, $^9$Be, $^{10}$B and $^{11}$B  through spallation of CNO nuclei with protons and alphas as well as the fusion of two $\alpha$-particles (in the case of Li isotopes). The resulting ``yields'' $y_6^{GCR}$ and $y_7^{GCR}$ (in \ms)  are time-dependent because they involve the composition of the ISM and of the GCR. The mono-isotopic $^9$Be is used as a test of the GCR composition, since observations of [Be/Fe] vs [Fe/H] in the past 30 years suggest that this composition has little changed over time \citep[and references therein]{Prantzos2012}.\footnote{With a few exceptions \citep[][and all subsequent works of those groups ]{Prantzos_1993, Lemoine_1998, Fields_1999,  Ramaty_2000}, most modellers of Li evolution consider the GCR Li component in a fairly simplified way, by assuming that it follows the behaviour of Be vs Fe, observed to be primary. This is not correct, since the Li isotopes are produced not only by spallation of CNO nuclei (as Be does) but also by fusion of alpha nuclei \citep[see e.g. Fig.~13 in ][]{Prantzos2012}. Moreover, Li is submitted to astration while Be considerably less so. But the impact of those effects on the evolution of the GCR component of Li is small, of the order of 10-20 \%.}

Nova eruptions are produced when H-rich material from a companion star is piled up on the surface of a white dwarf (composed of either C-O or O-Ne) and reaches ignition in degenerate conditions, at sufficiently high pressures. A fraction of the H-rich material is convectively mixed with the C-O-rich material of the white dwarf and H-burning takes place at temperatures of 200-300~MK through the hot CNO cycle. Nuclear reactions between the accreted $^3$He and $^4$He produce unstable $^7$Be that decays into $^7$Li with a half-life of $\sim$53.22~days. The fragility of  $^7$Be makes it almost impossible to survive at such high temperatures. But the rapid evolution and high mass loss rate during the nova eruption, allows $^7$Be to be transported away from the hot region and be ejected in the ISM.  

Although novae were proposed a long time ago as a key stellar source for Li production, the quantitative evaluation of its importance was hampered by two factors: the very uncertain Li yields of novae and the unknown DTD of those sources. Various empirical DTDs for novae -- with a nova rate proportional to the formation rate of whited dwarfs plus some time-delay -- were used in several papers \citep{DAntona_1991,Romano_1999,Romano2001,Grisoni2019}, while  \citet{Rukeya2017} adopted a population synthesis code to model the nova rate.  The very recent work of \cite{Kemp_phd} -- also presented in \citet{,Kemp_2022b,Kemp_2022a} -- provides a very promising theoretical framework for the nova DTD, and we adopt it here.

\begin{figure}
\includegraphics[width=1\linewidth]{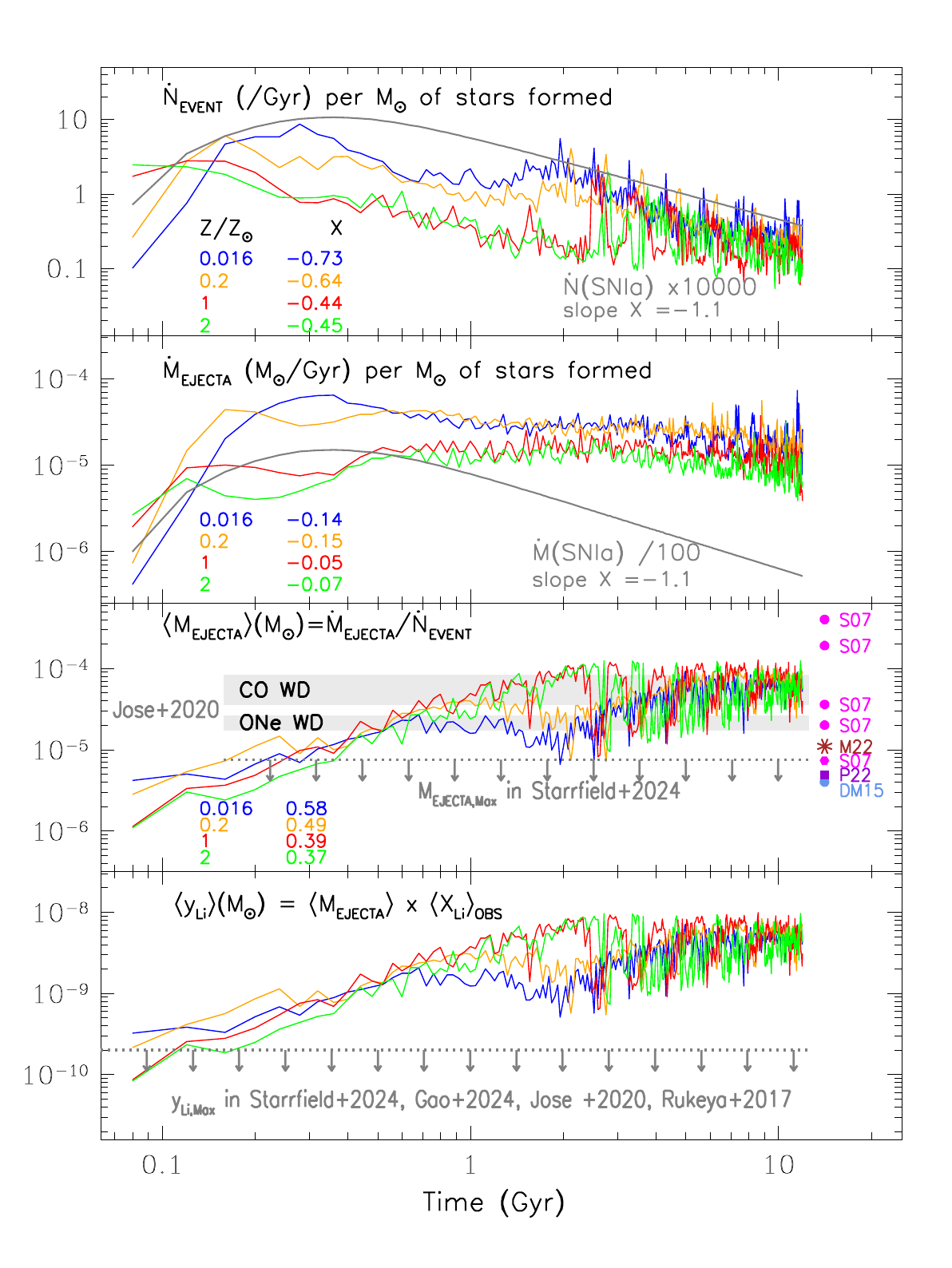}
\caption{DTDs for novae event rates $\dot{N}$  (\textit{top}) and ejecta mass rates $\dot{M}$  (\textit{second from top}) per 1 \ms \ of stars formed and for a few selected metallicities, adopted from \cite{Kemp_2022a}. The slopes X of power-law fits appear colour-coded for each metallicity. The grey curves display the corresponding DTDs for SN~Ia, with a single slope X=-1.1 above t=0.3~Gyr; the SN~Ia curves are multiplied by 10000 in the top panel and divided by 100 in the second panel. \textit{Third panel}: average ejecta mass of \cite{Kemp_2022b},  compared to the highest ejecta mass of the models of \cite{Starrfield2024} indicated by upper limits, to hydrodynamical models of \cite{Jose2020} within grey shaded areas and to available observations from S07 \citep{Schwarz2007}, M22 \citep{Molaro2022}, P22 \citep{Pandey2022}, and DM15 \citep{Das2015}. {\it Bottom panel}: Average Li yield, assuming the ejecta mass of the third panel and an average  Li mass fraction $\langle \rm X_{\rm Li}\rangle_{OBS}$ from observations (see Fig.~\ref{fig:nova_yields} and text); the grey arrows indicate the highest Li yield obtained in any of the cited studies.} 
\label{fig:dtd}
\end{figure}

In Fig.~\ref{fig:dtd}, we display the metallicity-dependent DTDs of \citet{Kemp_2022a} for both novae events and ejected masses because they have not the same DTDs. In that respect, the case of novae differs from the one of SN~Ia, in which only the DTD of SN~Ia rate is required, while the ejected mass is taken always to be approximately the Chandrasekhar mass. We show DTDs of nova events and ejecta for 1~\ms \ of stars formed and for a few selected metallicities Z in the two top panels, colour-coded with their metallicities. The curves are fitted with power-laws and the corresponding slopes appear also in the panels. Those slopes, and in particular the ones for the ejecta,  are systematically smaller -- in absolute value -- than the slope of x=-1.1 at late times for SN~Ia (with grey curves in both figures). This implies that nova production of Li will rise at late times more rapidly than Fe production from SN~Ia, and thus the [Li/Fe] ratio is expected to rise with metallicity in those times, as will be discussed in Sec.~\ref{subsec:results_multi-zone}.

In the third panel of Fig.~\ref{fig:dtd} we display the average mass of the nova ejecta, obtained by dividing the curves of the second panel by those of the first one. This average mass is time-dependent (increasing with time), because of the systematic difference in the slopes of the previous panels. The time-integrated average mass of the ejecta (over 12~Gyr) is $\langle \rm M_{EJECTA}\rangle \sim5\times10^{-5}$\ms \ per nova event and varies little between the various metallicities.  Compared to the results of recent hydrodynamical nova explosions from \cite{Starrfield2024}, these masses are several times higher, especially for the higher metallicities. In contrast, the 3D hydrodynamical models of \citet{Jose2020}, which appear within grey shaded areas in the figure, obtain values in the range of those provided in \cite{Kemp_2022b}. Those values are within the range of observations (indicated on the right of the third panel), which display a rather wide diversity.

The work of \cite{Kemp_2022b} adopts a novel approach in the study of the nova impact on GCE, by introducing the correspondence of ejecta DTDs to different regions of the parameter space of M$_{WD}$ vs $\dot{M}_{accretion}$. This allows them to map physics-dependent yields from hydro-simulations to the relevant ejecta rates of their own calculations. It turns out, however, that the Li yields of current nova models are notoriously insufficient to produce the meteoritic value of Li. For that reason, \citet{Kemp_2022b} adopt Li mass fractions from observations (see below) and they evaluate Li yields by multiplying those mass fractions with their own ejecta masses; we adopt that approach here. The obtained yields are provided in the bottom panel of Fig.~\ref{fig:dtd} and are considerably higher than those of the hydro models of \cite{Starrfield2024} or other recent studies \citep{Gao2024,Jose2020,Rukeya2017}. On the other hand, despite the high ejecta masses found in the new 3D hydro-simulations of \cite{Jose2020}, the authors find a serious decrease (by more than a factor of 50) in the Li amount with respect to their previous calculations.

At this point, it should be noted that current hydrodynamical models of nova explosions (1D, 2D or 3D) suffer from various shortcomings, due to the poorly understood mixing processes (shear, diffusion, Kelvin-Helmholtz instabilities) that operate in the interface between the core (CO or ONe WD) and the accreted envelope (H-rich). The mixing is responsible for the metallicity enhancement observed in the nova ejecta, and its modalities determine critically the amount of surviving $^7$Be \citep[see][]{Jose2020}.

\begin{figure}
\includegraphics[width=0.95\linewidth]{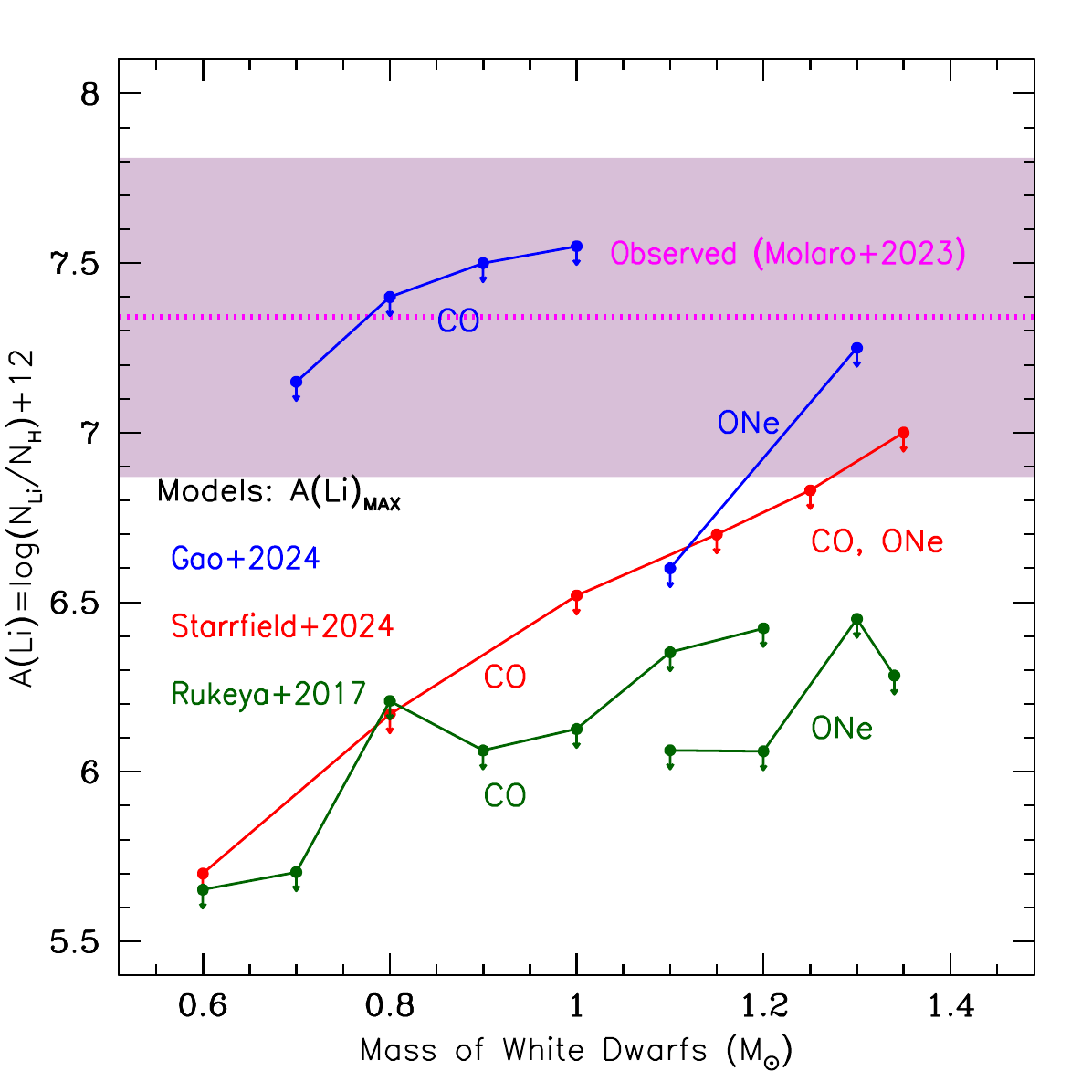}
\caption{Relative abundances of Li/H in nova explosions.  The shaded area shows the range of observed values and the red horizontal line is the average value, as reported in \citet{Molaro2022}.  Curves represent the highest values found in the calculations of \citet[][in red]{Starrfield2024}, \citet[][in green]{Rukeya2017} and \citet[][in blue]{Gao2024}, in explosions of CO and ONe white dwarfs of various masses. The results of \citet{Jose2020} are considerably below the scale of the figure.}
\label{fig:nova_yields}
\end{figure}

The discrepancy of current hydro-models and observations regarding the Li mass fraction in novae appears very clearly in Fig.~\ref{fig:nova_yields}. Only the most massive ONe white dwarf models of \cite{Starrfield2024} produce a Li abundance compatible with the lowest range of the observed values. A surprising exception comes from the very recent work of \citet{Gao2024} who find $^7$Be mass fractions from CO nova compatible with the observed ones. The authors claim that this is due  to the  high mass fraction of $^3$He assumed in their calculations \citep[an idea suggested in][]{Molaro2020} combined  to their treatment of element diffusion, while acknowledging that \citet{Denissenkov2021}  find the opposite effect, i.e. that an enhanced amount of $^3$He reduces the $^7$Be mass fraction. However, the ejecta masses in  the models of \citet{Gao2024} are quite small and so are their corresponding Li yields (see bottom panel in Fig. \ref{fig:dtd}).

In those conditions, it appears to us that the theoretical Li yields are  by far insufficient to explain the proto-solar Li value at present.
For that reason, we adopt here the mean $^7$Be abundance observed in novae, estimated by \citet{Molaro2022,Molaro2023} (dotted horizontal line in Fig.~\ref{fig:nova_yields}) as A($^7$Be)=$7.34\pm0.47$ which corresponds to Be (and, thereof,  Li) mass ratio over H of $X_\mathrm{Li}/X_\mathrm{H}=7\times10^\mathrm{A(Be)-12}=(1.53^{+2.99}_{-1.01})\times 10^{-4}$. To convert this ratio into the required Li mass fraction, the corresponding mass fraction of H is needed and this depends on the nova model. Nova modellers assume mixtures of white dwarf material (H-free) and accreted material (with solar H), of either 25\% - 75\% or 50\% - 50 \% \citep[e.g][]{Jose1998,Starrfield2024}, leading to $X_\mathrm{H}\approx0.5$ for the former case and $X_\mathrm{H}\approx0.35$ for the latter. The corresponding Li mass fractions in the ejecta are then $X_\mathrm{Li}=(7.7^{+14.9}_{-5.1})\times 10^{-5}$ and $X_\mathrm{Li}=(5.4^{+10.5}_{-3.5})\times 10^{-5}$. In order to calculate the Li production rate here, we use the former value multiplied by the nova ejecta DTDs (second panel in Fig.~\ref{fig:dtd}) after folding them with the SFR. On the other hand, the time-integrated average values of the ejecta reported previously ($5\times10^{-5}$ \ms) multiplied by the above values lead to an average value of the Li yield of a nova of $(3-4)\times10^{-9}$ \ms, 15-20 times higher than the maximum Li yields obtained in all recent studies \citep{Starrfield2024,Gao2024,Jose2020,Rukeya2017}, as indicated in the bottom panel of Fig. \ref{fig:dtd}.

\subsection{Results based on the one-zone model}
\label{subsec:results_one-zone}

In this paper, we focus on novae as the key stellar source of Li production. However, within the context of the one-zone model, we initially explore the possibility of AGB stars as a potential source of Li. This possibility has been studied in the past \citep[e.g.][]{Travaglio2001,Romano2001,Prantzos2012} and rejected, since the then available AGB yields were too small to explain the pre-solar (meteoritic) Li value. \citet{Cristallo2015} do not provide Li yields and we assess here briefly that point in the light of the metallicity- and mass-dependent AGB yields of Li from \citet{Karakas2016}. 

\begin{figure}
\includegraphics[width=1\linewidth]{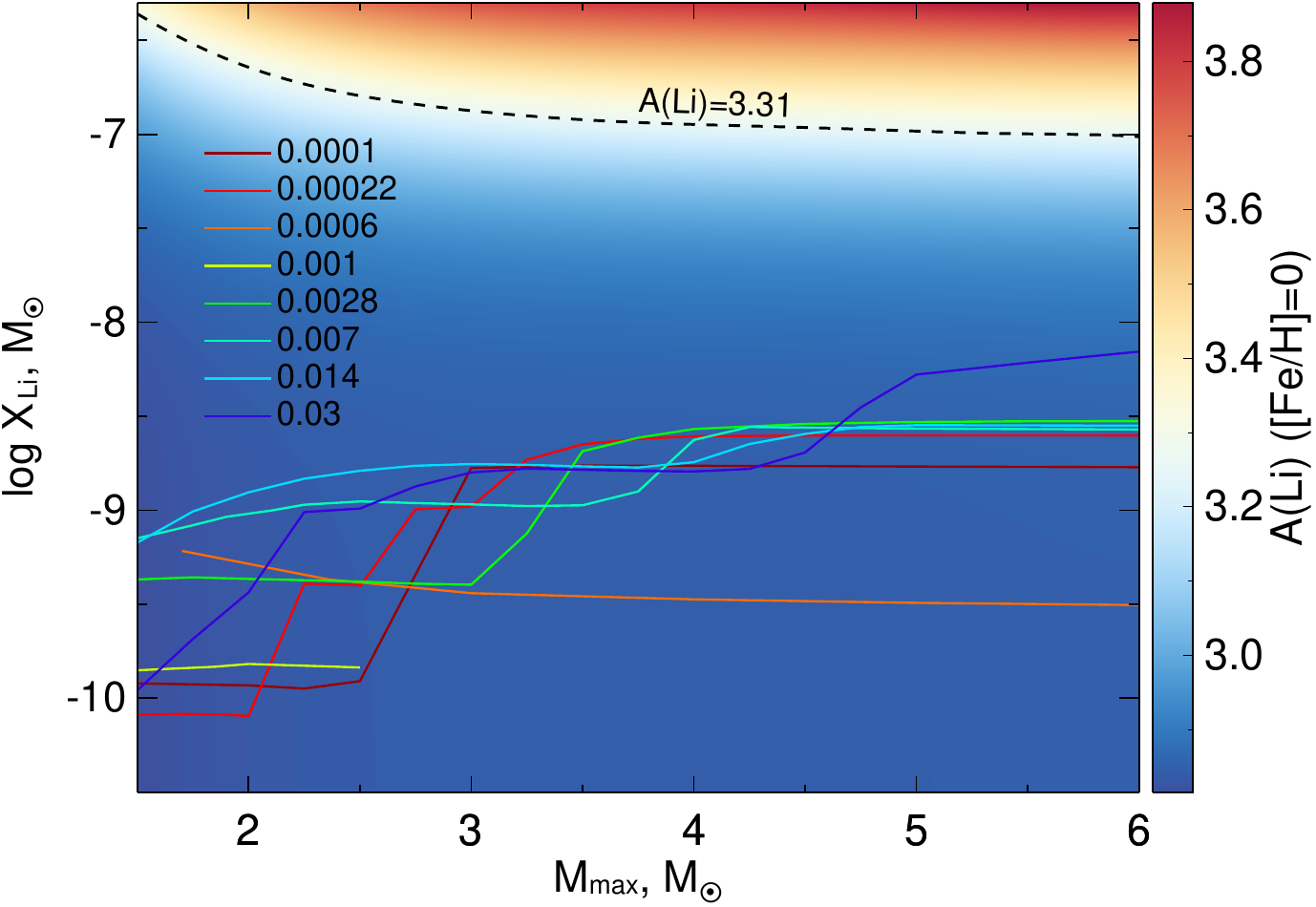}
\caption{Average Li yields of AGB stars in the range 1.5 to $\rm M_{max}$ as a function of $\rm M_{max}$ (according to Eq.~\ref{eq:AvLi_yield}) from \citet{Karakas2016}, colour-coded according to the initial stellar metallicity. They are compared to the average yield required to obtain the meteoritic Li value at [Fe/H]=0 (dashed curve).}
\label{fig:agb_yields}
\end{figure}

In our approach, we assume that AGB producers of Li cover the range of initial masses from 1.5~\ms \ to M$_\mathrm{max}$. We compute then models with different values of M$_\mathrm{max}$ and flat yields $X_\mathrm{Li}$ in the corresponding mass range. The value of A(Li) at solar metallicity and birth-time (4.56~Gyr ago) is shown colour-coded in Fig.~\ref{fig:agb_yields} as a function of both parameters, M$_\mathrm{max}$ and  $X_\mathrm{Li}$. The dashed line in the figure indicates the combinations of M$_\mathrm{max}$ and $X_\mathrm{Li}$ that provide the meteoritic value of A(Li)=3.31 at [Fe/H]=0. The required Li yields are slightly higher than 10$^{-7}$~\ms \  in the range from 1 \ms \ to M$_\mathrm{max}$.  Such high yields are found only in the supersolar-metallicity AGB stars of \citet{Ventura_2020} but only for masses higher than 7~\ms, while they are much smaller at lower masses. In fact, below 5 \ms \ stars display no Hot Bottom Burning and rather destroy their initial Li in those models (depending on the metallicity).

In the same figure, we show the average yields of AGBs with initial masses from 1.5~\ms \ to M$_\mathrm{max}$ and different metallicities according to \citet{Karakas2016}. For each value of M$_\mathrm{max}$, the average yield is computed as IMF-weighted mean:
\begin{equation}
\langle X_\mathrm{Li} \rangle = \frac {\int_{1.5 \mathrm M_{\odot}}^{\mathrm M_\mathrm{max}} X_\mathrm{Li}(\mathrm M) \mathrm M^{-\alpha} \mathrm{dM} }{  \int_{1.5 \mathrm M_{\odot}}^{\mathrm M_\mathrm{max}}  \mathrm M^{-\alpha} \mathrm{dM}    },
\label{eq:AvLi_yield}
\end{equation}

where $\alpha=2.3$ according to \citet{Kroupa2001}. The average yield of AGBs obtained that way is far below the one required to produce the meteoritic Li value at \feh=0, by almost 2 orders of magnitude on average. Thus,  AGB stars cannot be the main stellar source of Li, at least with presently available yields. In fact, the discrepancy in that respect is even larger than in \cite{Prantzos2012}, who had adopted older AGB yields.

\begin{figure}
\includegraphics[width=1\linewidth]{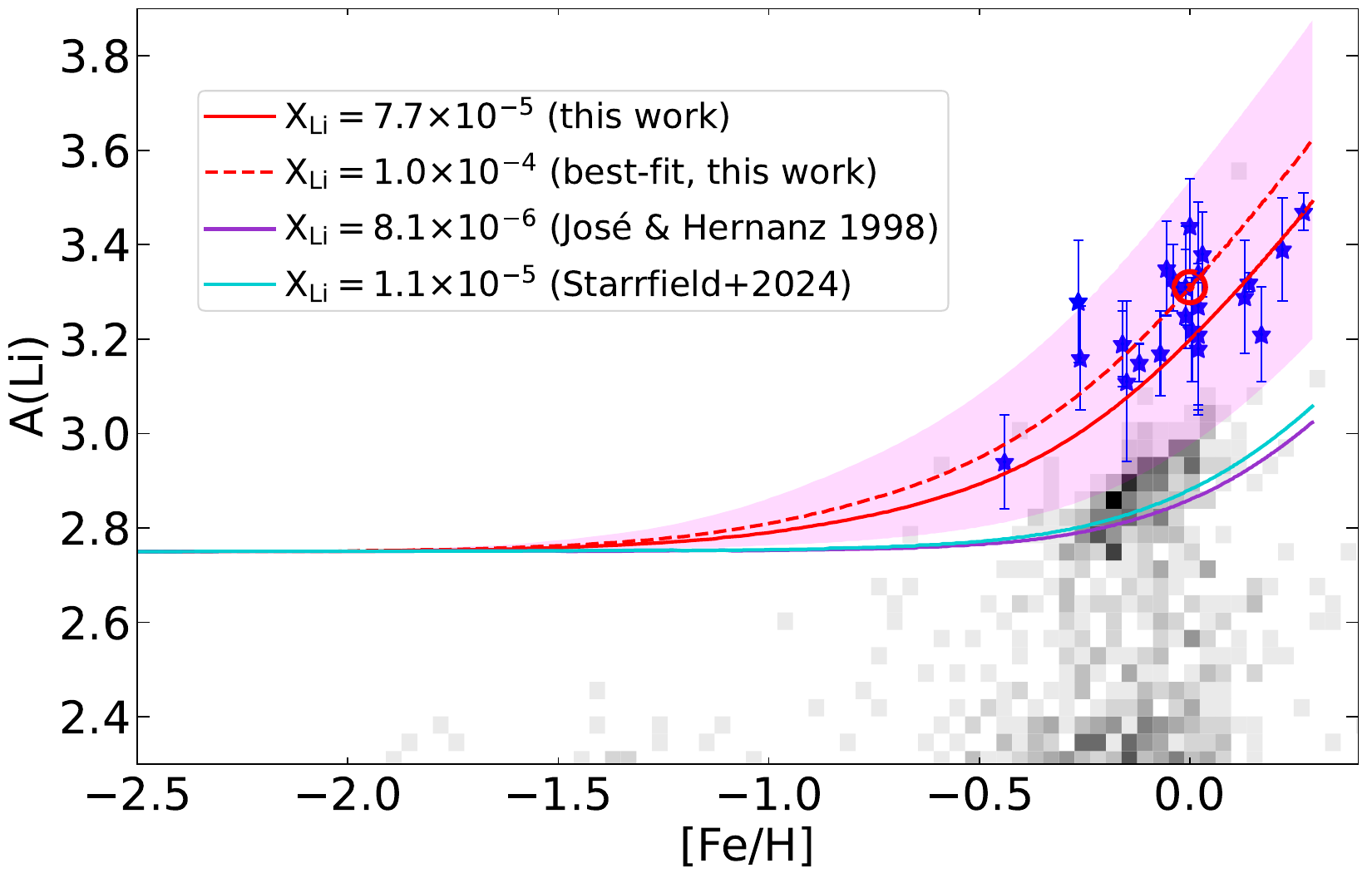}
\caption{Li abundance as a function of \feh \ with nova as a sole stellar source and with different mass fractions of Li in the nova ejecta. Our ``best-fit'' value (red dashed line) is slightly higher than the mean value of the observational analysis of \cite{Molaro2022} (red solid line), and the pinkish shaded area corresponds to the dispersion of $X_\mathrm{Li}$ around that mean value. The theoretical nova Li yields of \cite{Jose1998} and \cite{Starrfield2024} are also used (purple and cyan curves, respectively). The five-pointed stars show the abundance of the open clusters, while the underlying number density plot represents the GALAH sample. The solar symbol indicates the meteoritic value.} 
\label{fig:li_evol_1zone}
\end{figure}

Using the same one-zone GCE model, we studied the case of nova as a sole stellar source of Li in the Galaxy by using the metallicity dependent DTDs from \citet{Kemp_2022a} for nova rates of events and ejected masses (as shown in Fig.~\ref{fig:dtd}). From our modelling, we find (Fig.~\ref{fig:li_evol_1zone}) that the nova mass fraction that best fits the solar meteoritic value of Li is $X_\mathrm{Li}$=$1.0\times 10^{-4}$, which is consistent with the observational average estimate of Fig.~\ref{fig:nova_yields} within $\sim$$0.25\sigma$ of A($^7$Be). However, the large dispersion of A($^7$Be) gives room for variation in the mixing percentage, as indicated with the shaded area in Fig.~\ref{fig:li_evol_1zone}. The four curves are obtained with four different values of Be (Li) mass fraction: the best-fit value that we found, the mean value based on the estimation by \citet{Molaro2022} and 25\% mixing, and also the maximum possible values of the mass fraction from hydrodynamical simulations by \citet{Jose1998} and \citet{Starrfield2024}; the latter two are clearly insufficient to make the meteoritic Li value in the framework of our model \citep[see also][]{Kemp_2022b}.

\begin{figure}
\includegraphics[width=1\linewidth]{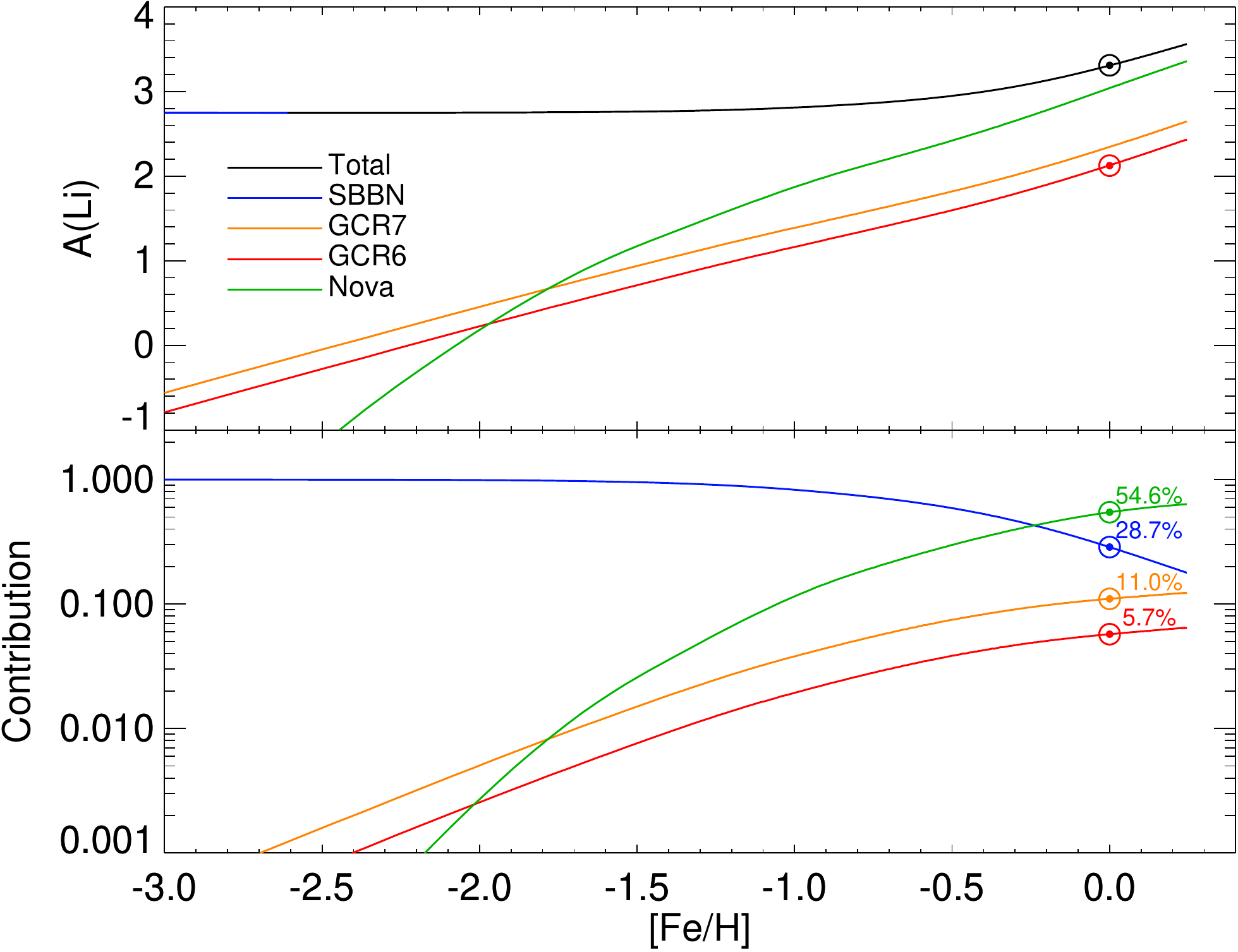}
\caption{Evolution of Li abundance (\textit{top}) and relative contributions (by mass) of different Li sources (\textit{bottom}) as a function of \feh. In the \textit{top} panel, the evolution of total abundance is shown in black, and the solar symbols indicate the solar meteoritic values for $^6$Li and $^7$Li. The other curves (in both panels) show the components from the BBN, novae, and GCR (the latter both for $^6$Li and $^7$Li). In the \textit{bottom} panel, the solar symbols indicate the contributions at solar metallicity.} 
\label{fig:li_contribution_1zone}
\end{figure}

In Fig.~\ref{fig:li_contribution_1zone} (top panel), one sees that GCR reproduce correctly the meteoritic abundance of $^6$Li a solar metallicity, a key test regarding the treatment of the GCR component. Following \cite{Prantzos2012}, we present the relative contributions of the various Li sources in Fig.~\ref{fig:li_contribution_1zone} as a function of time in the framework of this one-zone model. Those contributions correspond to the time-integrated masses 
of Li ejected during the evolution by each of its 3 different sources (BBN, GCR, and novae) and are calculated as follows:
\begin{equation}
C_{BBN}(t) = \int_0^t  \dot{m}_{INF}(t') X_7^{BBN} \ dt'     
\end{equation}
where $\dot{m}_{INF}$(t) (\ms/Gyr) is the rate of infall with primordial composition;
\begin{equation}
C_{GCR}(t) = \int_0^t  R_{SN}(t') y_7^{GCR}(t')+y_6^{GCR}(t')] \ dt'     
\end{equation}
where  R$_{SN}$(t) is the total supernova rate (CCSN + SN~Ia) per Gyr and  $y_6^{GCR}$ and $y_7^{GCR}$ (in \ms) are the (time and metallicity dependent) GCR ``yields'', as discussed in Sec.~\ref{subsec:nova_treatment};
\begin{equation}
C_{NOVA}(t) = \int_0^t  \dot{m}_{NOVA}(z(t'),t') \ \langle X_{\rm Li}^{NOVA}\rangle_{OBS} \  dt' 
\end{equation}
where  $\dot{m}_{NOVA}$(z(t),t) is the nova ejecta rate in \ms/Gyr, obtained by the convolution of the SFR and the metallicity-dependent nova ejecta DTDs of \cite{Kemp_2022a},  and 
$\langle X_7^{NOVA}\rangle$ is the average mass fraction of Li observed in novae, as discussed in Sec.~\ref{subsec:nova_treatment}.

During the early period of the Galactic evolution, primordial Li dominates the total Li abundance;  this Li is in the form of Li-7 and is contained both in the original gas and in the gas of primordial composition continuously accreted by the Galaxy. The contribution of GCR ($\sim$1/3 of that in the form of Li-6 and the other 2/3 as Li-7, by mass) reaches $\sim$0.1 \% of the total around [Fe/H]$\sim$-2.5, while the contribution of nova reaches that level around [Fe/H]$\sim$-2. The sum of those two exceeds the primordial contribution only around [Fe/H]$\sim$-0.5 when the metallicity was slightly above 1/3 solar. The relative contributions to the proto-solar (meteoritic) Li are approximately: 28\% from primordial nucleosynthesis, 17\% from GCR, and 55\% from nova. Thus, although the stellar component definitely dominates the abundance of Li at Sun's formation, the other two components almost match it. But 4.5~Gyr later (today), the stellar contribution of novae is increased by 10\% and the BBN contribution is decreased by a similar amount. Today, the nova component clearly dominates, having made $\sim$2/3 of the Li present in the Galaxy.

We note that in \citet{Prantzos2012}, the primordial contribution to Li in the proto-Sun was found to be smaller than here (12\% instead of 28\%), mainly because in that work the adopted primordial value of Li was A(Li)=2.6 instead of A(Li)=2.75 considered here and the meteoritic value A(Li)=3.4 instead of 3.3 here.

\section{Multi-zone chemical evolution models for the disc of the Milky Way}
\label{sec:multi-zone}

We extend our analysis to the study of the whole disc by using a multi-zone model, which offers more insight into the chemical evolution of the Galactic disc. Such models use more constraints, both global  (e.g., the present-day total mass of gas and stars, rates of star formation, infall, supernova, and nova, etc.) and local ones (e.g. radial profiles of gas, stars, and metals). These constraints result from the variation of various physical processes across the disc and enable a more detailed study of its history.

\begin{figure}[h]
\includegraphics[width=1\linewidth]{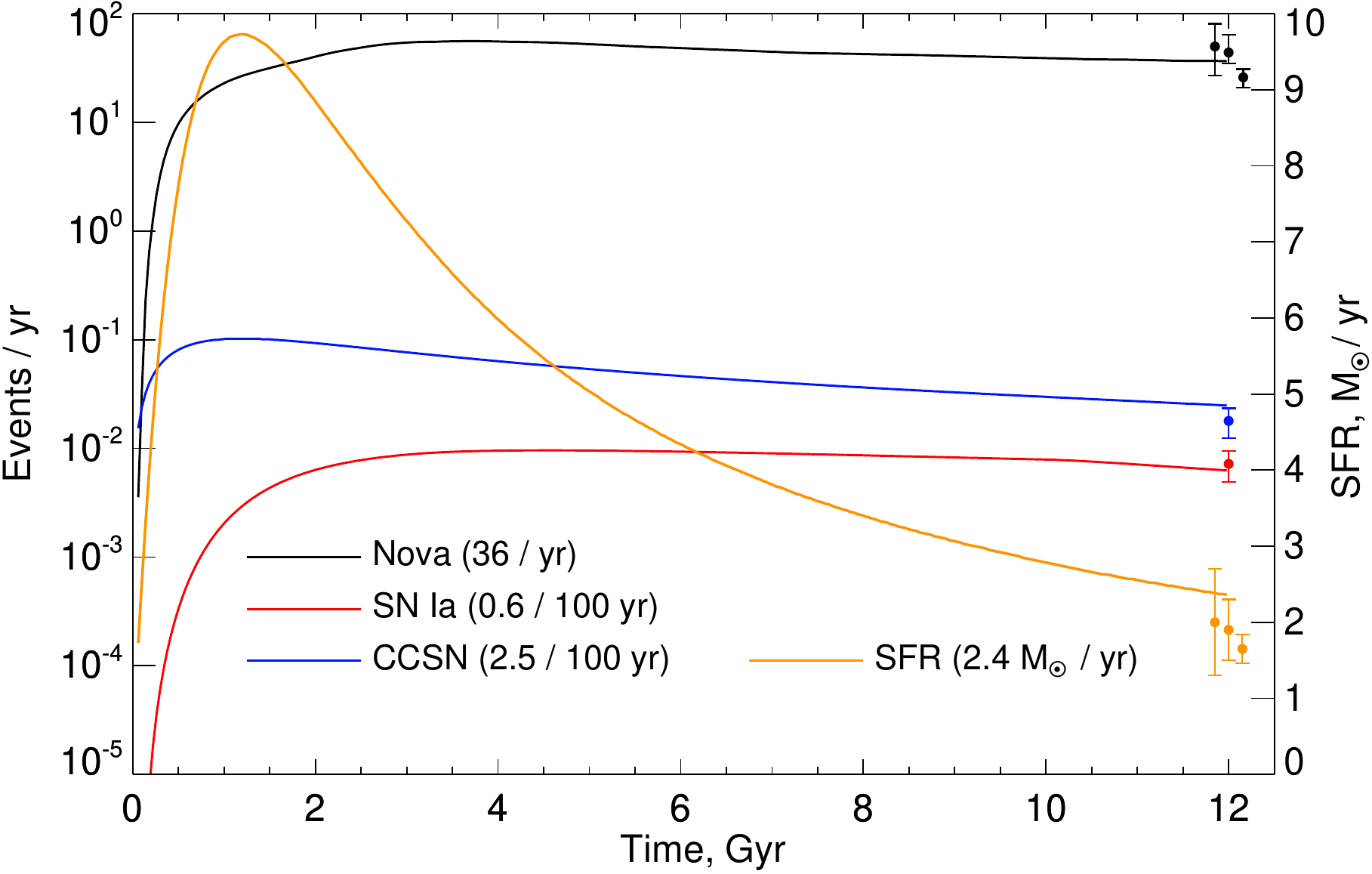}
\caption{Evolution of total rates of star formation (SFR, orange), novae (black), SN~Ia (red), and CCSN (blue) in the Milky Way predicted by our model. In the legend, the final model values (at 12~Gyr) are indicated. The SFR values are on the right axis (linear scale). Points with error bars show the present-day estimated values (see Sec.~\ref{subsec:results_multi-zone} for references).}
\label{fig:rate}
\end{figure}

We use the code described in the recent work of \cite{Prantzos2023}, which is an update of \cite{Kubryk2015a}. Here we present briefly its main features. The  Galactic disc is formed by the infall of primordial gas into the gravitational well of an evolving dark matter halo, with a final mass of $\sim$$10^{12}$\ms. The gas infall timescales are shorter in the inner regions, leading to the inside-out formation of the disc. Stars are formed from molecular gas, the amount of which is calculated from the amount of total gas through semi-empirical prescriptions. Stars undergo radial displacements due to epicyclic motions (``blurring'')  as well as variations in their guiding radius (``churning''). Radial migration is modelled as a diffusion phenomenon, with coefficients derived from N-body simulations.  As initially discussed in \citet{Kubryk2013}, radial migration influences not only ``passive tracers'' of chemical evolution (i.e., long-lived stars preserving the chemical composition of their birth gas in their photospheres), but also ``active agents'' of chemical evolution, namely long-lived nucleosynthesis sources,  such as SN~Ia producing iron, low-mass stars producing s-process elements and neutron-star mergers as producers of r-elements. Lithium is now added to that list since novae are considered here as its main producers. The stellar IMF and the nucleosynthesis part of the model are the same as in the one-zone model described in the previous section and we focus here only on novae as stellar sources of Li.

\subsection{Results of the multi-zone model}
\label{subsec:results_multi-zone}

Some global results of the multi-zone model are displayed in Fig.~\ref{fig:rate}. The total SFR (linear scale on the right) is rather high (almost 10~\ms/yr) for the first Gyr, being fuelled mostly by the gas of the inner disc, while it declines smoothly at late times to a value $\sim$2.4~\ms/yr, compatible with the observed ones  \citep[e.g.][]{Elia2022,Chomiuk2011,Licquia2015} reported the values of $2.0\pm0.7$, $1.9\pm0.4$, and $1.65\pm0.19$ respectively (all three values are shown in Fig.~\ref{fig:rate}). The same trend is followed by the CCSN rate (logarithmic scale on the left), while the SN~Ia rate stays almost constant during the whole evolution, because of the DTD of SN~Ia, as weighted by the SFR history of the various zones. In both cases, the final values of those rates are compatible with the present-day values of those quantities, reported in the figure ($\mathrm{R_{CCSN}}=1.79\pm0.55~\mathrm{yr^{-1}}$ \citep{Rozwadowska2021} and $\mathrm{R_{Ia}}=0.72\pm0.23~\mathrm{yr^{-1}}$ \citep{Maoz2014}).

Most importantly for our study, the total nova rate -- which is dominated by the inner regions and declines only by a factor of two in the last 8~Gyr --  reaches a final value of 36~novae~yr$^{-1}$, in fair agreement with the observationally inferred ones (e.g., \citet{Shafter2017}, \citet{De2021}, and \citet{Kawash2022} provide estimations of $\mathrm{R_{nova}}=50^{+31}_{-23}$, $44^{+20}_{-9}$, and $26\pm5$~novae~yr$^{-1}$, respectively). It is the first time -- to our knowledge -- that the nova event DTD of \citet{Kemp_2022a} is used in a multi-zone model tailored to reproduce the properties of the Milky Way disc \citep{Prantzos2023} and reproduces correctly the present-day nova rate of our Galaxy (within observational uncertainties). This is encouraging, both regarding the work presented for the one-zone model (in which the present-day theoretical nova rate could not be checked by local observations) and what follows in this section.

\begin{figure}
\includegraphics[width=1\linewidth]{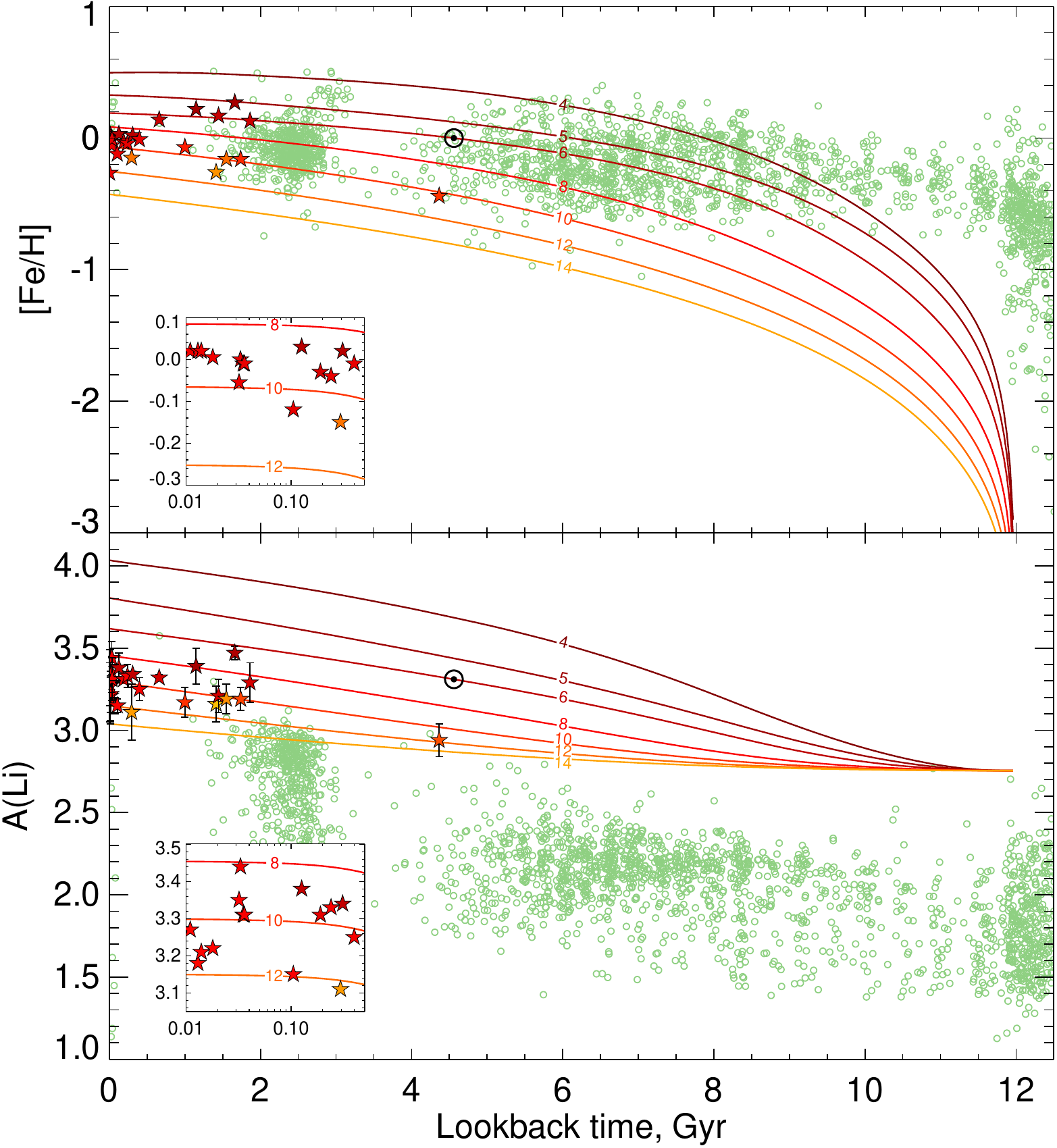}
\caption{\feh \ (\textit{top}) and A(Li) (\textit{bottom}) as a function of age for the GALAH~DR3 stars (green circles) and open clusters (five-pointed stars colour-coded according to their Galactocentric radius). The solid curves (colour-coded the same way as the open cluster symbols) show the model evolution of A(Li) in different radial zones (4, 5, 6, 8, 10, 12, and 14~kpc, from top to bottom). The insets in both panels focus on the very last 0.5~Gyr of the Galactic evolution. The Sun is formed at Galactocentric radius $R_{\rm GC}$ 6~kpc, 4.56~Gyr ago.} 
\label{fig:feh_and_li_vs_age}
\end{figure}

Although the Galactic rate of nova events appears to be $\sim$ constant in time in Fig.~\ref{fig:rate}, this is not true for the Galactic rate of the mass of nova ejecta: the slope of their DTD with time (Fig.~\ref{fig:dtd}) is considerably smaller in absolute value than the one of the nova event DTD. As a result, the nova ejecta rate increases with time, especially in the inner Galactic zones, leading to a continuous increase of the Li abundance, in contrast to the outer zones (see Fig.~\ref{fig:feh_and_li_vs_age}).

In Sec.~\ref{subsec:results_one-zone} we adjusted the Li mass fraction of novae to reproduce the value of Li/H in the model 4.56~Gyr ago (observed meteoritic value). In the multi-zone model, the same should be done for the zone of Galactocentric distance $R_{GC}$=8~kpc, in the absence of radial migration. However, several observational indices suggest that the Sun was not formed in that zone but rather in the inner disc and subsequently migrated to its current position. 

Indeed, observations show that the gaseous abundances of several elements within $\pm$1~kpc from the Sun are solar to within $\pm$0.04~dex \citep{Cartledge2006}. Also, observations of young B-stars both in the field and in the nearby star-forming region of Orion \citep{Nieva2012}, find solar abundances with small dispersion. Finally,  recent observations of \cite{Ritchey2023} find clear evidence that the dispersion in the metallicities of neutral interstellar clouds in the solar neighbourhood is small ($\pm$0.10~dex) and only slightly larger than the typical measurement uncertainties. Taking into account the observed abundance gradient, the above results suggest that the Sun migrated to its present-day position from its birthplace, located a couple of kpc inwards in the disc \citep[e.g.,][]{Nieva2012}. Theoretical arguments for such a displacement, invoking radial migration, have been proposed \citep[e.g.,][]{Wielen1996, Minchev2014,Kubryk2015a}. As discussed in \cite{Prantzos2023}, we find in the framework of this model that [Fe/H]=0 is obtained 4.56~Gyr ago in the gas of the zone at Galactocentric radius $R_{GC}\sim$6~kpc. We then use the results for that zone for the calibration of the nova Li mass fraction.

The results appear in Fig.~\ref{fig:feh_and_li_vs_age}  where the evolution of the gas abundances for Fe (top) and Li (bottom) is presented for zones at various galactocentric distances. For the zone at 6~kpc, we find a ``best-fit'' value $X_\mathrm{Li}=7.2\times10^{-5}$, i.e. about 2/3 of the one determined in the one-zone model but still well within the range of observed values reported for novae in \citet{Molaro2022}. Taking into account the uncertainties involved in our model (including, among others, the actual extent of the radial migration of novae in the Galaxy) as well as in the observations, we find that the result is quite satisfactory and strengthens the confidence in the role of novae as major stellar contributors to Li enrichment in the Milky Way.

An interesting feature is that in the inner disc, the Fe abundance practically stops increasing at late times, while the abundance of Li increases continuously to very high values. The former is due to the large amounts of gas returned from low-mass stars, which are abundantly produced in the first few Gyr from the intense star formation in those regions; this gas, released at late times, is metal-poor and dilutes the local metallicity. 
In contrast, Li is less diluted, because the gas returned from low-mass stars is Li-free, but the infalling gas contains primordial Li.
Moreover, the DTD of nova ejecta declines more slowly with time than the SN~Ia DTD (see second panel in Fig.~\ref{fig:dtd}); as a result, the strong early activity of those inner regions produces a larger ratio of nova/SN~Ia ejecta at late times, enhancing more and more the Li abundance compared to the one of Fe.

Another interesting feature, related to the above, is that for the same metallicity, Li is more abundant in the outer zones than in the inner ones. This, rather counter-intuitive, result was first obtained in  \citet{Prantzos2017} and is due to the fact that the metallicity [Fe/H] rises very rapidly early on (within a few Gyr) in the inner zones, because of the intense star formation and the large contribution of CCSN. During that period, the Li abundance rises slowly, because a small fraction of novae have the time to release their Li. In contrast, the outer zones reach similar levels of [Fe/H] much later, at a time when a larger fraction of nova have operated. Interestingly, the opposite happens for products of massive stars (alpha-elements), which have very short DTDs, as we discuss below.

In Fig.~\ref{fig:feh_and_li_vs_age}, we have also plotted the corresponding data for main sequence stars from GALAH (see Sec.~\ref{subsec:selection} for details on the sample selection) and open clusters \citep[OC, blue asterisks,][]{Romano2021}. With the exception of the highest look-back time (most uncertain ages), the [Fe/H] values of GALAH correspond well to the evolution of Fe in our model zones of 6-10~kpc for the last 6~Gyr and to zones inside 6~kpc for older ages. However, the vast majority of the Li values of GALAH at all ages (and even more so for stars older than 2.5~Gyr) are far below the BBN value, a clear signature of Li depletion in stellar envelopes. This conclusion is independent of the formation zone of the GALAH main sequence stars, which are all observed locally.

In the framework of our model, the four OCs with the highest [Fe/H] and ages 1-2~Gyr are formed in the region of $R_{GC}$=5 to 6~kpc. However, the observed highest Li abundance of those OCs is substantially smaller than the model abundances in those zones and ages: it corresponds to stars formed outside $R_{GC}$=8~kpc. This discrepancy is again attributed to the depletion of Li on the surfaces of those stars, formed in the inner disc, in the last couple of Gyr. 

The inserts in Fig.~\ref{fig:feh_and_li_vs_age} zoom in on the youngest ages 0.01 to 0.05~Gyr, providing a different picture: the abundances of Fe and Li$_{max}$ of those clusters correspond to the model predictions for the zones 8-10~kpc. Those very young OCs were formed 
locally -- having little time to migrate -- and for the same reason they had little Li depletion in their envelopes.

\begin{figure}
\includegraphics[width=1\linewidth]{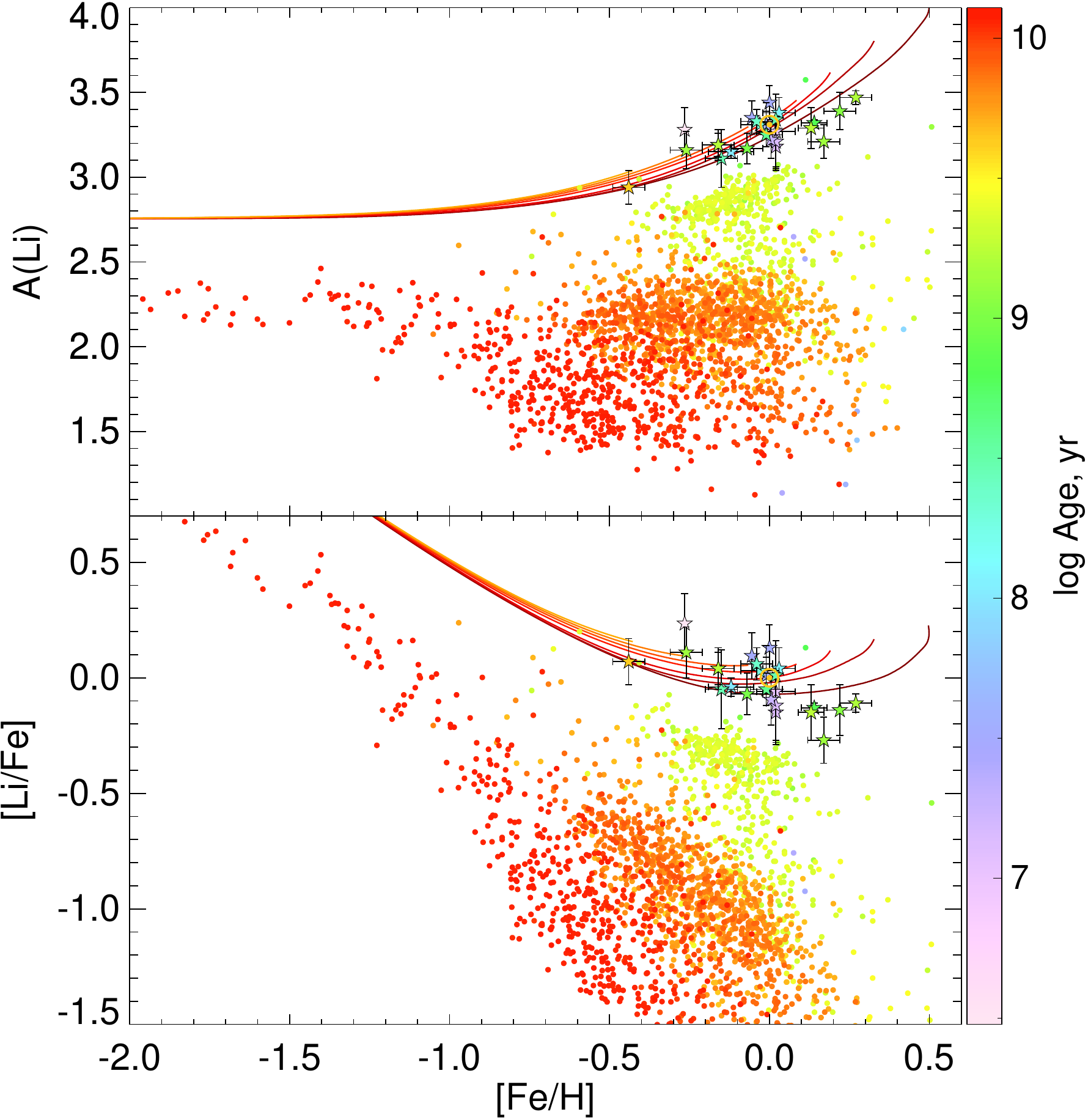}
\caption{A(Li) and [Li/Fe] as a function of \feh \ (\textit{top} and \textit{bottom} panels respectively) for GALAH~DR3 stars (circles) and open clusters from \citet{Romano2021} (five-pointed stars). The age of stars and open clusters is colour-coded. In both panels, the solid curves show model A(Li) in different radial zones (4, 5, 6, 8, 10, 12, 14~kpc from bottom to top, colours from dark red to orange as in Fig.~\ref{fig:feh_and_li_vs_age}).}
\label{fig:li_vs_feh_and_afe}
\end{figure}

Fig.~\ref{fig:li_vs_feh_and_afe} (top) displays the commonly studied A(Li) vs [Fe/H] diagram for the various model zones. Because of the high initial A(Li), the various curves evolve very closely to each other and start exceeding the primordial value around [Fe/H]$\sim$-1, but they display no significant divergence  (higher than an observational uncertainty of 0.1~dex) up to the solar value. As discussed before, at a given metallicity, it is the outer zones that have the higher Li abundance in their gas. Above solar metallicity, the rise of A(Li) concerns only the inner zones, because at 8~kpc, gas metallicity stops at [Fe/H]$\sim$0.1~dex, (comparable to one of the young local stars and ISM). In those inner zones, the abundance of Li continues increasing, more rapidly than the one of Fe which is affected by the aforementioned dilution effect from the Fe-poor ejecta of low-mass stars.

Compared to the observations, there is a fair agreement of the various curves (within observational uncertainties) with the open cluster data when [Fe/H]$>$-0.5~dex. In the highest metallicities, up to [Fe/H]=0.3~dex, the data seem to be well explained by the evolution of the innermost zones ($<$4~kpc), with no need to assume any Li depletion in the stellar envelopes. However, this interpretation is incorrect. In Fig.~\ref{fig:feh_and_li_vs_age}, the age and [Fe/H] of those super-solar metallicity clusters correspond to their formation at $R_{GC}\sim5$~kpc, while the corresponding observed A(Li) is lower than predicted for that zone at that time and it is thus attributed to depletion in the stellar envelope. This idea was already invoked in \cite{Guiglion2019}, who used different data (from AMBRE, lacking stellar ages and hot enough stars) and different models (with initial Li from the ``Spite plateau'', no radial migration and different prescriptions for nova DTD). This interpretation of Li abundances at supersolar metallicities is consistent with the ages of those clusters, around 1-2~Gyr, which is consistent with sizeable Li depletion in their stars. In contrast, no depletion is required in order to interpret the data around solar [Fe/H], since those clusters are very young, less than 0.2~Gyr old \citep[see also][]{Dantas_2022}. This is consistent with recent stellar evolution models \citep[e.g.][]{Dumont2021b,Dumont2021a}.

In addition to the open cluster data, Fig.~\ref{fig:li_vs_feh_and_afe} also includes data for field stars from the GALAH survey. These field stars show a larger scatter in A(Li) (top panel) and [Li/Fe] (bottom panel, discussed below) compared to the open clusters due to factors such as large span in age (from 25~Myr to 12.8~Gyr) and \teff \ (meaning that the sample also contains G- and K-type stars that deplete Li much faster than their F-type counterparts). While the GALAH data are not directly comparable to the models, they can thus be seen as providing useful observational limits.

The bottom panel of Fig.~\ref{fig:li_vs_feh_and_afe}
provides another, original, view, of the Li vs metallicity relation. [Li/Fe] decreases with increasing [Fe/H] as does [$\alpha$/Fe] (not shown). However, the reason in the case of Li is not the different DTDs of the corresponding sources (short DTD for $\alpha$-elements from CCSN vs long DTD for Fe from SN~Ia) but simply the initial abundances (primordial for Li vs zero for Fe, so that [Li/Fe] can only decrease with time). Around solar metallicity, [Li/Fe] flattens because the main sources of the two elements (novae for Li and SN~Ia for Fe) have similar DTDs. Finally, at the highest metallicities of the innermost zones [Li/Fe] rises up again because of the differential dilution discussed before (higher for Fe than for Li). 
 
\begin{figure}
\includegraphics[width=1\linewidth]{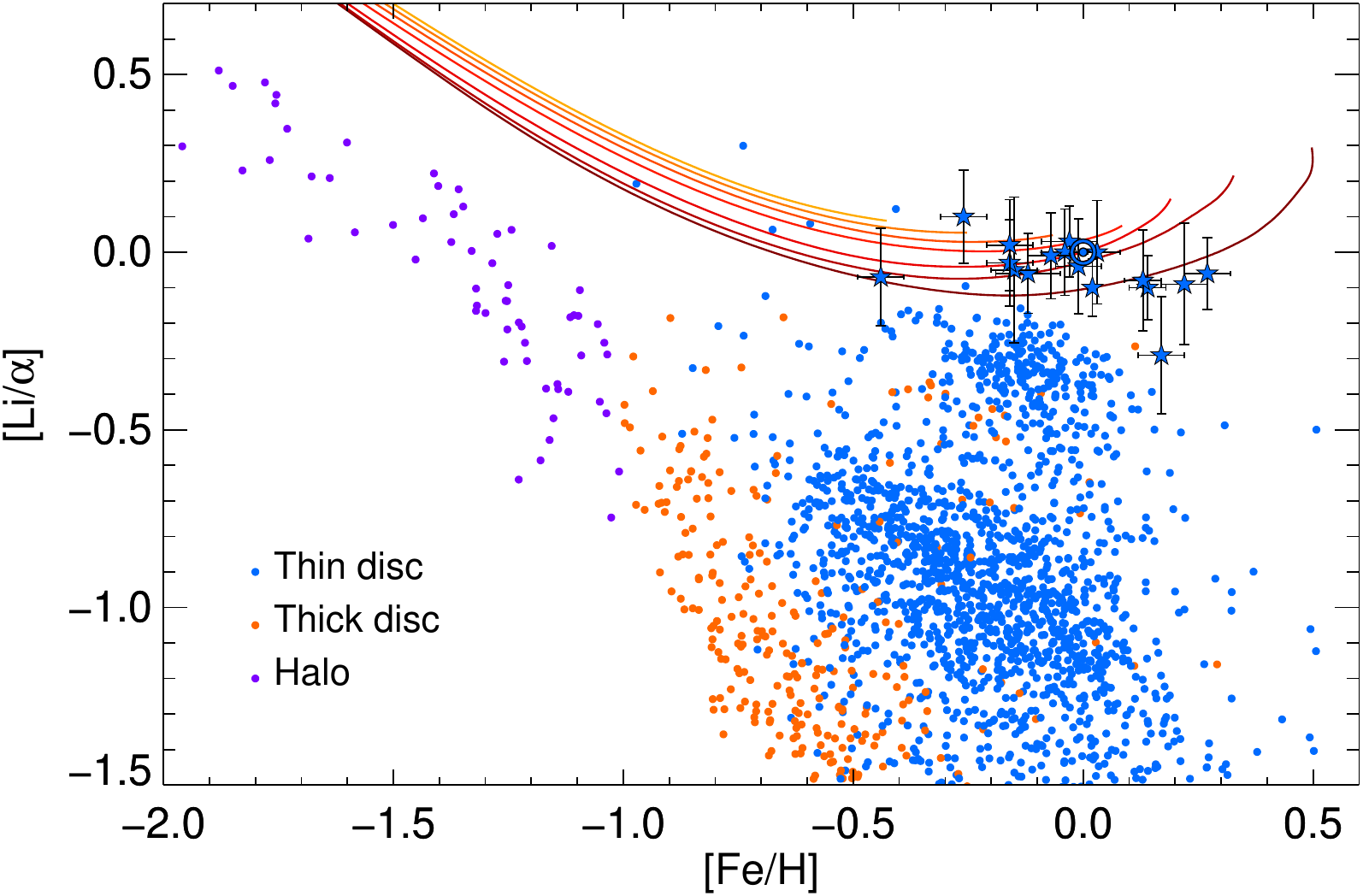}
\caption{ [Li/$\alpha$] as a function of \feh from GALAH, with thin disc stars (blue), thick disc (orange), and halo (purple).  We select the latter as those with \feh$\leq$-1~dex regardless of their \afe, while we distinguish between thin and thick discs on the basis of their [$\alpha$/Fe] by the Gaussian Mixture Model (GMM). Model curves are colour-coded as in Fig.~\ref{fig:li_vs_feh_and_afe}}
\label{fig:lia_vs_feh}
\end{figure}

Furthermore, as already mentioned, at a given metallicity [Li/Fe] is higher in the outer zones with their younger (low [$\alpha$/Fe]) population than in the inner ones with their older stars (high [$\alpha$/Fe] population); this behaviour is opposite to the one of [$\alpha$/Fe] vs metallicity: at a given metallicity, the [$\alpha$/Fe] ratio is lower in an inner than an outer zone, as it is well known from models of the chemical evolution of the Milky Way \citep{Prantzos2023}. This difference, due to the vastly different DTDs of the sources of those elements (very short for $\alpha$-elements, quite long for Li with the adopted nova ejecta DTDs), is clearly illustrated in Fig.~\ref{fig:lia_vs_feh}, where the [Li/$\alpha$] ratio is plotted vs \feh. The model curves are distinguishable now even at low metallicities, in contrast to the situation for [Li/Fe] (bottom panel in Fig. \ref{fig:li_vs_feh_and_afe}). 

The observed double-branch behaviour of [$\alpha$/Fe] vs metallicity (with the high \afe \ values at a given \feh \ corresponding to the thick disc and the low ones to the thin disc)  is interpreted in our model through the combined effects of radial migration, secular evolution and different DTDs for $\alpha$ elements (produced by massive stars) and Fe (mostly produced by SN~Ia), as discussed in detail in \citet{Kubryk2015a,Prantzos2023}: stars from the inner, rapidly evolving regions (with high \afe \ values) are found today in the local disc which has undergone a slower evolution (leading to lower \afe \ values for the same metallicities)\footnote{We note that there exist other interpretations of the double-branch behaviour of [$\alpha$/Fe] vs metallicity, invoking not secular evolution, but a paucity in the star formation for a few Gyr affecting partially or totally the Galactic disc \citep[e.g.][and references therein]{Grisoni2019}}. A similar question arises obviously for Li since the DTDs of novae (Li producers) are more extended than those of SN~Ia (Fe producers). However, the observational situation is not yet clear. Li abundances in the local thick disc are definitely lower than those of the thin disc, but the upper Li envelope is very poorly defined in the latter case. Starting from the ``Spite-plateau'', \citet{Ramirez_2012} report a flat A(Li) versus \feh \ behaviour up to solar metallicity, while \citet{Guiglion_2016} and \citet{Fu_2018} find an increase of the upper Li envelope with metallicity and \citet{Bensby_2018,Bensby_2020} find a decrease. The various difficulties plaguing the separation of the discs (by age, kinematics, or chemistry, respectively) and the subsequent evaluation of the upper Li envelopes are summarized in \cite{Smiljanic_2020}.
 
From the theoretical side, \citet{Prantzos2017} made the first attempt to quantify the effect with a multi-zone model with radial migration. They found that, if the high Li abundance of BBN is adopted along with a long-lived source of Li (as here), the upper envelope of the thick disc stars should lie above the observations up to a metallicity \feh$\sim$-0.5, and this is also the case here  (compare their Fig. 3 with the top panel of our Fig. \ref{fig:li_vs_feh_and_afe}). This implies that up to that metallicity, it is Li depletion that determines the observed upper Li envelope, requiring appropriate stellar physics for its interpretation \citep[][]{Borisov2024}.

In conclusion, the interpretation of the upper envelope of the Li abundance vs metallicity requires: from the observational side stars hot enough  \citep[preferentially above the Li-dip, see discussion in][]{Charbonnel2021} and an estimate of their ages; and from the theoretical side, models of Li depletion in stellar envelopes for various metallicities and multi-zone models of Galactic evolution with radial migration.

\subsection{Li across the Galactic disc}
\label{subsec:open_clusters}

\begin{figure}
\includegraphics[width=1\linewidth]{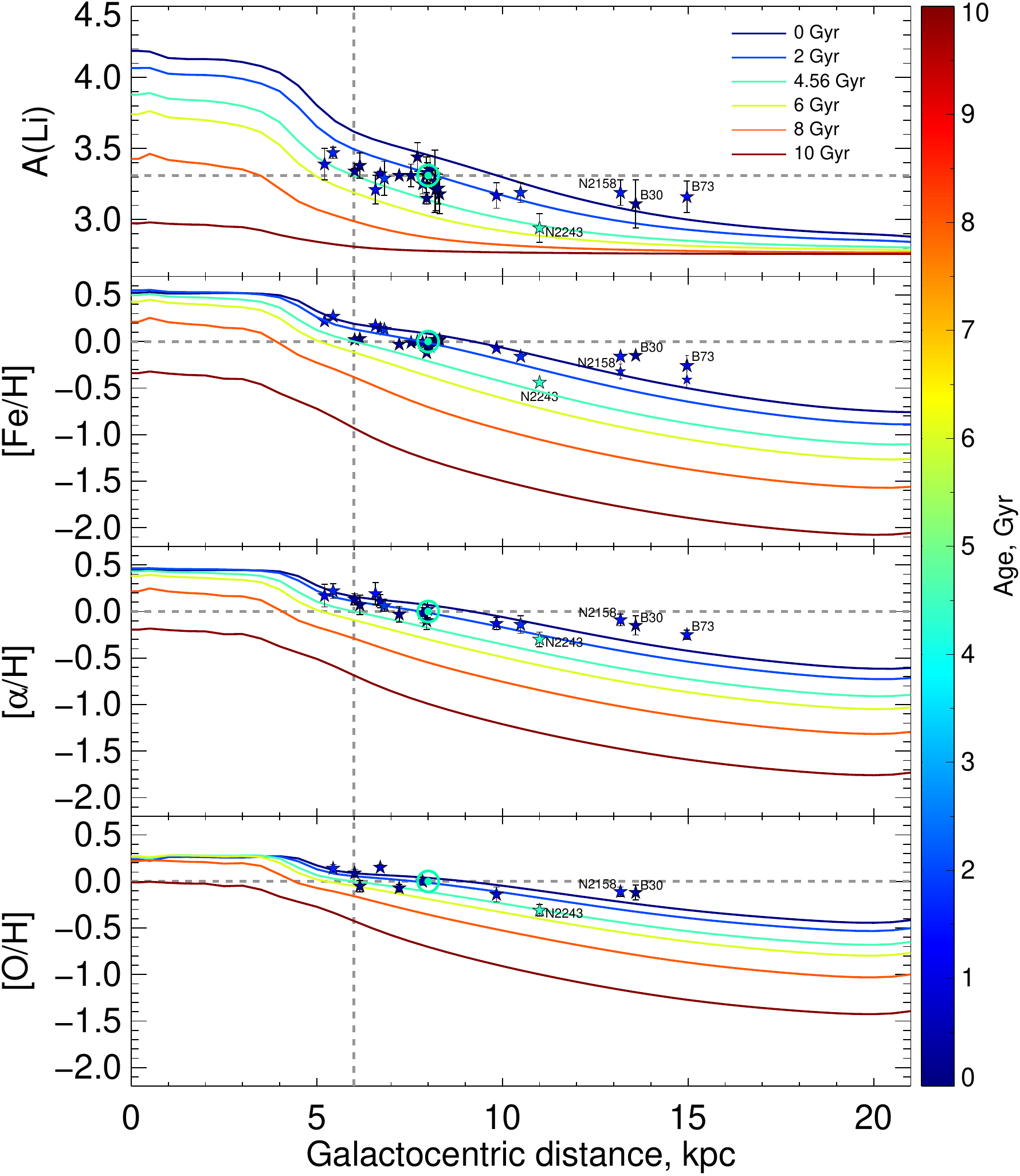}
\caption{Radial profiles of A(Li), \feh, [$\alpha$/H], and [O/H] in the galactic gas of our model. Different solid lines show the profiles at different ages (lookback times) and have colours according to their age. The five-pointed stars show the position and abundances of open clusters according to \citet{Romano2021} with age colour-coded. For Berkeley~73 and NGC~2158, we show additional \feh \ estimations from \citet{Netopil2016} with smaller five-pointed stars. The grey dashed lines show the solar birth radius and abundances, while the solar symbol shows the present-day location.} 
\label{fig:gradient_feh_li}
\end{figure}

Figure~\ref{fig:gradient_feh_li} shows the predicted radial profiles of A(Li), \feh, [$\alpha$/H], and [O/H] in the Milky Way gas at different stages of the evolution in our model (every 2~Gyr in age). The final  (present-day) gradient of [Fe/H] is found to be d\feh/d$R_{GC}$ $\sim$-0.07~dex/kpc in the range $5<R_{GC}~(\rm kpc)<15$ \citep{Prantzos2023}, which is slightly steeper than values estimated from Cepheid data: $-0.06$~dex/kpc \citep{Genovali2014} or  $-~0.05$~dex/kpc \citep{Luck2018}. 
We notice here that the abundance profile of Fe (or any other element) is not necessarily characterized by a unique slope over the whole radial range.  The shape of the abundance profile, flat in the innermost 4~kpc, which is steeply declining in the intermediate region 4 $<$ R/kpc $<$ 14 and is less steep in the outer disc, resembles the one obtained in \cite{Magrini2009}, who adopted very different prescriptions for the star formation, infall, and SN~Ia rates. A break of the slope in the inner disc is also obtained in other studies, e.g. \cite{Minchev2014}.

For Li, we find a present-day gradient of d[Li/H]/d$R_{GC}$ $\sim-0.06$~dex/kpc, slightly steeper than in \cite{Prantzos2012} who found $-0.05$~dex/kpc and slightly flatter than in \citet{Romano2021} with -0.07~dex/kpc. Those models, however, used different prescriptions for the nova rate and had no radial migration. We note that, in contrast to the Fe gradient which becomes flatter with time (steeper with age) because of the inside-out disc formation, the Li gradient starts quite flat (because of the initial Li from BBN) and steepens considerably at late times because of the adopted nova DTDs.

The, now indisputable,  impact of radial migration on the evolution of the Galactic disc, makes it difficult to infer many of its past properties. This concerns, in particular, the determination of the birth radius of disc stars from observations, and therefore the determination of the evolution of the abundance gradient in the ISM (i.e. at the birthplace of stars that are currently observed locally). \cite{Minchev2018} attempted to estimate the birth radii by projecting locally observed mono-age stars in radius along hypothesized age gradients and using the metallicity distributions of those mono-age populations (from HARPS-GTO survey) to constrain the result. Recently, \cite{Zhang2023} used the result of \cite{Minchev2018} to infer the birthplace of stars and open clusters (OCs) with measured Li abundances from observations with the multi-subject optical fibre facility FLAMES. They found that the highest Li abundances of local and young field stars and OCs show no sign of depletion in the stellar envelopes, while stars born in the inner disc are older on average and are Li-depleted.

Subsequent studies on the evolution of the [Fe/H] gradient at birthplace from local observations showed that the gradient does not evolve monotonically \citep{LuMinchev_2022} and displays two recent fluctuations,  attributed to recent star formation episodes \citep{Ratcliffe_2023}; moreover, the limitations and drawbacks of the method have been underlined already in \cite{Minchev2018}  and in more detail in \citet{Lu_Buck_2022}. We shall thus refrain here from projecting our sample of field stars to their birth radii. 

In Fig.~\ref{fig:gradient_feh_li}, we overplot on our model results data for young OCs from \citet{Romano2021}. We compute [$\alpha$/H] as a error-weighted mean of [Mg/H] and [O/H] based on values provided by \citet{VanderSwaelmen2023}. For those clusters that do not have measurements of [O/H], we adopt [$\alpha$/H]=[Mg/H]. We do not show \feh \ uncertainties in this figure since they are lower than the size of the symbols. 

As has already been discussed, Li is subject to depletion in stellar interiors. With this in mind, we see a relatively good agreement between the data and the model predictions for most OCs. This is the case for the very young OCs in the solar neighbourhood, with \feh, [$\alpha$/H], and [O/H] values corresponding to the model predictions in the last Gyr of the evolution: their A(Li) is slightly lower than the model prediction, as expected if they had known some Li depletion in their atmosphere. For stars in the inner disc, the agreement between the model and observations is fairly satisfactory for the heavy elements and corresponds to the model metallicity gradient; however, the required amount of Li depletion (taking into account the model predictions) is larger than for the local OCs, as expected from their higher age.

The oldest OC, NGC~2243 with the age of $\sim$4.4~Gyr, is at the same time the most iron-poor and Li-poor one. Its values of \feh, [O/H] and A(Li) correspond well to the radius and age of the Galactic model, with no need for radial migration or Li depletion in its atmosphere. On the other hand, from the stellar evolutionary models that we computed and described in \citet{Borisov2024}, we expect Li depletion from the initial abundance of at least $\Delta$A(Li)=0.25-0.30~dex at this age for a star with the composition of the cluster and \teff \ range of the stars that were used to compute the OC parameters (6064-6314~K, see \citet{Romano2021} for details). Taking into account the observational uncertainty of $\pm\sim0.1$~dex in A(Li), we think that the properties of this cluster are well described by the model.

In the outer disc, the situation appears less satisfactory, since all observed abundances are higher than the model predictions. This discrepancy may be due to an underestimate of the rate of evolution in those outer regions because of the model prescriptions.  Alternatively, it may be due to the possibility that those OCs have migrated radially from their birthplace in the inner disc where metallicity was higher than in their current position today \citep[see, e.g.][and references therein]{Dantas_2023}. Below, we scrutinize each one of the three cases of OCs in the outer disc.

We see a significant overabundance, both in \feh \ and $\alpha$-elements, for the two outer OCs, NGC~2158 (Age$_\mathrm{N2158}$=1.55~Gyr) and Berkeley~73 (Age$_\mathrm{B73}$=1.4~Gyr). The GES values of metallicity are \feh$_\mathrm{N2158}$=-0.16 and \feh$_\mathrm{B73}$=-0.26~dex which is $\sim0.2-0.25$~dex higher than the model's predictions. The difficulty is alleviated if we adopt the lowest values from the literature (\feh$_\mathrm{N2158}$=-0.32$\pm$0.08 \citep{Netopil2016} and \feh$_\mathrm{B73}$=-0.41$^{+0.26}_{-0.09}$ \citep{Perren2022}; those values are shown with smaller stars in Fig.~\ref{fig:gradient_feh_li}). For Li, one should consider a higher initial value than the one appearing in the figure, to account for the depletion expected by the cluster age. However, the uncertainties in A(Li), combined with the possibility of some amount of radial displacement (a couple of kpc is reasonably expected during the $\sim$1.5~Gyr ages of those OCs) again alleviate the discrepancy between the model and observations.

Finally, the very young open cluster Berkeley~30 (Age$_\mathrm{B30}$=0.3~Gyr),  shows an overabundance of \feh \ and [$\alpha$/H] with respect to the model. This could indicate an overestimation of the metallicity for this cluster or highlight the limitations of the model in accounting for specific local conditions or processes affecting individual clusters. According to the models of \citet{Dumont2021a,Dumont2021b}, the expected Li depletion at this age is very small, $\Delta$A(Li)$\approx$0.07~dex. Thus, despite the challenges with [Fe/H] and [$\alpha$/H], the cluster's A(Li) is consistent with the model predictions well within the observational uncertainties. 

Overall, we think that the position of the studied OCs in the 5-dimensional  phase space of ($R_{GC}$, age, \feh, A(Li) and [O/H]) are well described by the model, especially if on considers all the observational uncertainties related to Li determination, nicely resumed in the discussion of \citet{Romano2021}. Also, it should be noted, however, that for the youngest OCs, the observed [Fe/H] values appear flatter than predicted by the model. This discrepancy could suggest that the model overestimates the steepness of the gradient for young populations. However, given the small number of OCs in this regime and the uncertainties involved in our model, we refrain from drawing firm conclusions.

\section{Summary}
\label{sec:conclusion}

In this work, we reassess the evolution of Li in the Galaxy, focusing on the role of novae. Although that source was proposed a long time ago as a key stellar source for Li production, the quantitative evaluation of its importance was hampered by two factors: the very uncertain Li yields of novae and the unknown DTD of those sources. Recent hydrodynamical models do not help with the former problem, finding very low Li mass fractions in the nova ejecta and thus insufficient Li yields, more than 20 times lower than required by simple theoretical expectations. This includes the surprising recent results of \citet{Gao2024}, who find high Li mass fractions but very low ejecta masses and thus low Li yields. In contrast, observations of $^7$Be abundances in nova explosions in the past ten years find sufficiently high values for the Li mass fraction in the nova ejecta. As for the nova DTDs,  the recent work of  \citet{Kemp_2022a}  provides a very promising theoretical framework (see Sec.~\ref{subsec:nova_treatment} and Fig.~\ref{fig:dtd}).

We first use a well-tested one-zone model of GCE to explore the impact of two potentially important stellar sources of Li,  namely AGB stars and novae. For the former source, we adopted the mass and metallicity dependent yields of stars with mass 1.5-8 \ms \ from \citet{Karakas2016}, who find Li production by Hot-Bottom Burning in stars more massive than $\sim 4$\ms.  Our analysis shows that AGB stars fall short (by a large factor) of reproducing the Li meteoritic abundance at the Sun's formation, 4.5~Gyr ago. Unless the Li production is seriously underestimated in current AGB calculations, those stars cannot be considered as an important Li source, as already discussed in several analogous studies in the past \citep{Travaglio2001,Romano2001,Prantzos2012}.

In contrast, our investigation of Li production from novae has yielded interesting findings. Our simple one-zone model shows that using the \citet{Kemp_2022a} DTD and within the bounds of Li mass fraction dispersion from nova observations, nova can reproduce the Li meteoritic abundance. This confirms previous findings, using similar Li yields, but different nova rates and different one-zone models of GCE \citep{Cescutti2019,Grisoni2019} and, of course, the results of \cite{Kemp_2022b} with the same nova DTDs. In the framework of our one-zone model, we find that the typical mass fraction in nova ejecta should be $X_\mathrm{Li}\sim$10$^{-4}$, i.e. rather on the upper range of current nova observations.

We find that the primordial contribution from BBN dominates the total Li abundance until a metallicity [Fe/H]$\sim$-0.5. The relative contributions to the proto-solar (meteoritic) Li are found to be approximately: 28\% from BBN, 17\% from GCR, and 55\% from nova; although the stellar component dominates the abundance of Li at Sun’s formation, the other two components almost match it. But today, the stellar contribution of novae  clearly dominates,
having made $\sim$2/3 of the Li present in the Galaxy. 

We then extend our analysis by using a multi-zone model,  presented in \cite{Prantzos2023}, which takes into account the radial migration of stars and various long-lived nucleosynthesis sources, including novae. 
A crucial test regarding the nova DTDs of \citet{Kemp_2022a} is provided by the current nova rate, found to be R$_{\rm Novae}\sim$36/yr in our model, in fair agreement with current observations.

In the framework of our multi-zone model, we find that the Sun was born about 2~kpc inwards from its current position, namely at galactocentric radius $R_{GC}\sim$6~kpc. The nova rate in that zone includes novae with progenitors locally formed, but also with progenitors which migrated from other Galactic regions. As a result, the mass fraction of Li in nova ejecta required to produce the meteoritic Li at $R_{GC}$=6~kpc and t$_{lookback}$=4.56~Gyr ago is found to be $\sim7\times10^{-5}$, close to the average range of observationally inferred values in \citet{Molaro2023}.
 
We find large differences in Fe and Li abundances between the various zones at any given time (Fig.~\ref{fig:feh_and_li_vs_age}). However, for a given metallicity, Li abundances vary little between the various zones (Fig.~\ref{fig:li_vs_feh_and_afe}) and are at the limits of observational uncertainties. This is due to the interplay between various factors: high early star formation activity in the inner zones vs low activity in the outer ones and different DTDs for the Fe ejected by SN~Ia and the Li ejected by novae, the latter DTD being more extended in time than the former. This is at the origin of a counterintuitive result, first obtained in \cite{Prantzos2017}, namely that at the same metallicity, Li is more abundant in the outer zones than in the inner ones, if its source is sufficiently long-lived.

We find that the most metal-rich OCs, with metallicities \feh=0.1-0.3~dex are formed in the region 5-6~kpc (Fig.~\ref{fig:feh_and_li_vs_age} top) and that their stars have suffered modest Li depletion of 0.2-0.3~dex during the 2-4~Gyr of their evolution. In general, taking into account the observational uncertainties, we find that the position of the studied OCs in the 5-dimensional  phase space of ($R_{GC}$, age, \feh, A(Li) and [O/H]) are well described by the model. 

Despite the success of the theoretical framework for nova provided recently by \cite{Kemp_phd}, it remains to be seen how Li observations can help to constrain the proposed nova ejecta DTDs, which differ substantially from the nova event DTDs. The latter are constrained by the observed Galactic nova rate, but the former require a good knowledge of the upper envelope of Li observations at various radii and a good understanding of the Li depletion in stars of various masses, ages and metallicities. Current (APOGEE, GALAH, RAVE, LAMOST) and forthcoming (4MOST, WEAVE)  surveys will contribute on the observational side \citep[see][and references therein]{Nepal2023}. In our recent work on Li depletion in stars of halo metallicities \citep{Borisov2024} we found excellent agreement between model predictions and observed Li trends in post-turnoff stars in the globular cluster NGC~6752. This opens perspectives for refining the estimates of initial Li abundance in metal-rich globular clusters (e.g., 47~Tuc and NGC~6496), thus providing additional constraints in the intermediate metallicity range between Pop~II and Pop~I stars. This integrated approach would lead to a deeper understanding of Li evolution from primordial nucleosynthesis to present-day observations.

\begin{acknowledgements}

We are grateful to Dr.~Alex Kemp for providing us with the set of his metallicity-dependent DTDs of nova events and ejecta; also to Jordi Jos\'{e} and Paolo Molaro, as well as to the referee Gabriele Cescutti, for a careful reading of the manuscript and very useful comments and suggestions that improved considerably the final text. We acknowledge support from the Swiss National Science Foundation (SNF; Project 200021-212160, PI CC).
\end{acknowledgements}

\bibliographystyle{aa}
\bibliography{gal_evol}

\end{document}